\documentclass[aps,pra,reprint]{revtex4-1}
\usepackage{amsmath}
\usepackage{amsfonts}
\usepackage{amssymb}
\usepackage{graphicx}
\usepackage{color}
\usepackage{hyperref}

\newcommand \commentout[1] {}

\newcommand \ket[1] {|{#1}\rangle}
\newcommand \bra[1] {\langle {#1}|}

\newcommand \nucidxA {I}
\newcommand \nucidxB {J}
\newcommand \nucidxC {K}

\newcommand \cartidxA {\alpha}
\newcommand \cartidxB {\beta}
\newcommand \cartidxC {\gamma}

\newcommand \mass {m}
\newcommand \charge {q}
\newcommand \TotMass {m_{\text{tot}}}

\newcommand \PES {v}
\newcommand \GVP {\boldsymbol{\chi}}
\newcommand \GVPcomp {\chi}
\newcommand \BerryTensor {\boldsymbol{\Xi}}
\newcommand \BerryTensorComp {\Xi}
\newcommand \BerryCurv {\boldsymbol{\Omega}^{\mathrm{int}}}
\newcommand \BerryCurvComp {\Omega^{\mathrm{int}}}

\newcommand \TotRot {\boldsymbol{\Omega}}
\newcommand \TotRotComp {\Omega}
\newcommand \mwTotRot {\widetilde{\boldsymbol{\Omega}}}

\newcommand \Hessian {\mathbf{H}}
\newcommand \HessianComp {H}
\newcommand \mwHessian {\widetilde{\mathbf{H}}}

\newcommand \TotRotGrad {\boldsymbol{\Lambda}}
\newcommand \TotRotGradComp {\Lambda}

\newcommand \DisplVec {\boldsymbol{\eta}}

\newcommand \DisplDeriv {\boldsymbol{\gamma}}

\newcommand \Inertia {\mathcal{I}}
\newcommand \QuadMom {\mathcal{Q}}

\newcommand \SkewSymEncode[1] {[{#1}]_{\times}}

\begin{document}

\title{Molecular vibrations in the presence of velocity-dependent forces}

\author{Erik I. Tellgren}

\email{erik.tellgren@kjemi.uio.no}
\affiliation{
Hylleraas Centre for Quantum Molecular Sciences, Department of Chemistry, University of Oslo, P.O.~Box 1033 Blindern, N-0315 Oslo, Norway}

\author{Tanner Culpitt}

\affiliation{
Hylleraas Centre for Quantum Molecular Sciences, Department of Chemistry, University of Oslo, P.O.~Box 1033 Blindern, N-0315 Oslo, Norway}

\author{Laurens Peters}

\affiliation{
Hylleraas Centre for Quantum Molecular Sciences, Department of Chemistry, University of Oslo, P.O.~Box 1033 Blindern, N-0315 Oslo, Norway}

\author{Trygve Helgaker}

\affiliation{
Hylleraas Centre for Quantum Molecular Sciences, Department of Chemistry, University of Oslo, P.O.~Box 1033 Blindern, N-0315 Oslo, Norway}

\begin{abstract}
A semiclassical theory of small oscillations is developed for nuclei that are subject to velocity-dependent forces in addition to the usual interatomic forces. When the velocity-dependent forces are due to a strong magnetic field, novel effects arise -- for example, the coupling of vibrational, rotational, and translational modes. The theory is first developed using Newtonian mechanics and we provide a simple quantification of the coupling between these types of modes. We also discuss the mathematical structure of the problem, which turns out to be a quadratic eigenvalue problem rather than a standard eigenvalue problem. The theory is then re-derived using the Hamiltonian formalism, which brings additional insight, including a close analogy to the quantum-mechanical treatment of the problem. Finally, we provide numerical examples for the H$_2$, HT, and HCN molecules in a strong magnetic field.
\end{abstract}

\maketitle 

\section{Introduction} \label{intro}

Many aspects of theoretical chemistry can be understood in terms of the semiclassical notion of a potential energy surface. Most importantly, equilibrium structures correspond to potential energy minima, while the curvature in different directions determines the molecular vibrational frequencies. These frequencies, in turn, are important for infrared spectroscopy and thermodynamical properties.

This simple picture derived from the potential energy surface needs to be modified when the nuclei are subject to velocity-dependent forces: There is then a distinction between the curvature of the potential energy and dynamical frequencies squared, as  noted in a previous work~\cite{TELLGREN_PCCP14_9492}. We are here interested the velocity dependence that enters in the form of the Lorentz force due to an external magnetic field. A sufficiently strong magnetic field has profound effects on the electronic structure and the potential energy surface. Chemical bonding is affected~\cite{LANGE_S337_327,TELLGREN_PCCP14_9492} and even many otherwise unbound one-electron ions become bound in a strong field~\cite{TURBINER_PR424_309}. Moreover, the center-of-mass motion becomes coupled to the internal motion and the Born--Oppenheimer approximation becomes less accurate~\cite{JOHNSON_RMP55_109,SCHMELCHER_IJQC40_371,SCHMELCHER_PRA38_6066,SCHMELCHER_JPB21_445}. Most such effects are beyond the scope of the present work. However, there is an important dynamical screening effect wherein the electrons partially shield the nuclei from the Lorentz forces due to the external field. A consistent semiclassical picture that takes this screening into account is that the electrons give rise not only to a Born--Oppenheimer scalar potential, as in standard treatments, but also a Born--Oppenheimer vector potential. The latter turns out to be the geometric vector potential associated with the Berry phase~\cite{YIN_JCP100_8125,CERESOLI_PRB75_161101,CULPITT_JCP155_024104,CULPITT_JCP156_044121,PETERS_JCP155_024105,MONZEL_JCP157_054106}.

In Sec.~\ref{secTheory}, we show how velocity-dependent forces modify the standard eigenvalue problem $(\mathbf{H}-\omega^2\mathbf{M})\hat{\DisplVec} = \mathbf{0}$, from which the mass matrix $\mathbf{M}$ and molecular Hessian $\Hessian$ determine vibrational frequencies $\omega$ and modes $\hat{\DisplVec}$. The modified problem contains a rotational term that couple translations, rotations, and vibrations. Because of this term, the modified problem is no longer a standard eigenvalue problem, but rather what is known as a quadratic eigenvalue problem. In Sec.~\ref{secHamiltonian}, the theory is considered in the Hamiltonian formalism, yielding a new formulation and a close analogy to quantum excitation energies. In Sec.~\ref{secResults}, we provide several numerical illustrations of physical effects that arise due to the new term.

\section{Theory} \label{secTheory}

Let $\mathbf{B}(\mathbf{s}) = \boldsymbol \nabla\times\mathbf{A}(\mathbf{s})$ denote the external magnetic field and $\mathbf{A}(\mathbf{s})$ its magnetic vector potential at position $\mathbf s$.
We adopt a semiclassical picture where the electrons are fully quantum mechanical and the electronic state $\psi(\mathbf{r};\mathbf{R})$ depends parametrically on the classical nuclear positions $\mathbf{R}$. Here, $\mathbf{r}=(\mathbf{r}_1,\ldots,\mathbf{r}_n)$ collectively denotes all electron coordinates and $\mathbf{R}=(\mathbf{R}_1,\ldots,\mathbf{R}_N)$ all nuclear coordinates. The Born--Oppenheimer scalar potential, or the potential energy surface, is given by
\begin{equation}
  \PES(\mathbf{R}) = \sum_{\nucidxA<\nucidxB} \frac{\charge_{\nucidxA} \charge_{\nucidxB}}{|\mathbf{R}_{\nucidxA}-\mathbf{R}_{\nucidxB}|} + \bra{\psi} H_{\mathrm{el}} \ket{\psi},
\end{equation}
where $\charge_{\nucidxA}$ is the nuclear charge and $H_\text{el}$ is the electronic Hamiltonian. The electronic Hamiltonian $H_\text{el}$ and wave function $\psi$ both depend parametrically on $\mathbf R$ but for ease of notation this dependence is suppressed.
The Born--Oppenheimer vector potential, or the geometric vector potential, is given by
\begin{equation}
  \GVP_{\nucidxA}(\mathbf{R})= -\mathrm i \int \psi^* \frac{\partial \psi}{\partial \mathbf{R}_{\nucidxA}} \mathrm d\mathbf{r}.
\end{equation}
Note that $\PES(\mathbf{R})$ and $\GVP_{\nucidxA}(\mathbf{R})$ depend on all nuclear coordinates jointly and the latter on an additional nuclear index. The sensitivity of the electronic state to geometric perturbations can be quantified by the tensor
\begin{equation}
  \BerryTensor_{\nucidxA\nucidxB}(\mathbf{R}) = \int \frac{\partial \psi}{\partial \mathbf{R}_{\nucidxA}} \left( \frac{\partial \psi}{\partial \mathbf{R}_{\nucidxB}} \right)^{\dagger} \mathrm d\mathbf{r}.
\end{equation}
The Berry curvature is defined in terms of $\GVP_{\nucidxA}$ as
\begin{equation}
  \label{eqBerryCurvDef}
  \BerryCurvComp_{\nucidxA\cartidxA,\nucidxB\cartidxB} = \frac{\partial \GVPcomp_{\nucidxA\cartidxA}}{\partial R_{\nucidxB\cartidxB}} - \frac{\partial \GVPcomp_{\nucidxB\cartidxB}}{\partial R_{\nucidxA\cartidxA}} = -i\BerryTensorComp_{\nucidxA\cartidxA,\nucidxB\cartidxB} + i\BerryTensorComp_{\nucidxB\cartidxB,\nucidxA\cartidxA},
\end{equation}
and coincides with the anti-symmetric part of the tensor $\BerryTensor$ that quantifies the sensitivity to rotational perturbations of the $3N$ dimensional vector $\mathbf{R}$. (The middle expression gives rise to second-order derivatives of $\psi$, which can be shown to cancel~\cite{CULPITT_JCP155_024104}.) For a single nucleus ($N=1$), the Berry curvature contains exactly the same elements as the curl $\nabla\times\GVP_{\nucidxA=1}$, only arranged as a skew-symmetric matrix rather than as a vector. For a system of several nuclei, it is a multi-dimensional generalization of the three-dimensional curl.

Newton's equation of motion, supplemented by the Berry screening force, now takes the form
\begin{equation}
  \label{eqNewtonEOM}
  \mass_{\nucidxA} \ddot{\mathbf{R}}_{\nucidxA} = -\frac{\partial \PES(\mathbf{R})}{\partial \mathbf{R}_{\nucidxA}} + \charge_{\nucidxA} \dot{\mathbf{R}}_{\nucidxA}\times\mathbf{B}(\mathbf{R}_{\nucidxA}) + \sum_{\nucidxB} \BerryCurv_{\nucidxA\nucidxB}(\mathbf{R}) \dot{\mathbf{R}}_{\nucidxB},
\end{equation}
where $\mass_{\nucidxA}$ is the mass of nucleus $I$. The Lorentz force has a more specific form than the Berry screening force; we therefore absorb the former into the latter. In component form, the Lorentz force is given by $F_{\mathrm{L};\nucidxA\cartidxA} = \charge_{\nucidxA} \epsilon_{\cartidxA\cartidxB\cartidxC} \dot{R}_{\nucidxA\cartidxB} B_{\cartidxC}(\mathbf{R}_{\nucidxA})$, where $\epsilon_{\cartidxA\cartidxB\cartidxC}$ is the Levi-Civita symbol and summation over $\cartidxB,\cartidxC$ is implied. Defining
\begin{equation}
  \label{eqTotRotDef}
  \TotRotComp_{\nucidxA\cartidxA,\nucidxB\cartidxB}(\mathbf{R}) = q_{\nucidxA}\delta_{\nucidxA\nucidxB} \, \epsilon_{\cartidxA\cartidxB\cartidxC}  \, B_{\cartidxC}(\mathbf{R}) + \BerryCurvComp_{\nucidxA\cartidxA,\nucidxB\cartidxB}(\mathbf{R})
\end{equation}
and the mass matrix $M_{\nucidxA\cartidxA,\nucidxB\cartidxB} = q_{\cartidxA\cartidxB} \,\delta_{\nucidxA\nucidxB} \, \mass_{\nucidxA}$, the equation of motion is compactly expressed in terms of $3N\times 3N$ matrices as
\begin{equation}
    \mathbf{M} \ddot{\mathbf{R}} = -\frac{\partial \PES(\mathbf{R})}{\partial \mathbf{R}} + \TotRot(\mathbf{R}) \, \dot{\mathbf{R}}.
    \label{eqmotion}
\end{equation}
We note that the Berry curvature is a real-valued antisymmetric $3N \times 3N$ matrix, $\boldsymbol \Omega^\text T = - \boldsymbol \Omega$.

\subsection{Rigid motion and translational-rotational coupling}

Before specializing the theory to small oscillations, we consider a different special case. Suppose that the motion consists purely of rigid translations and rigid rotations but no vibrations. All velocities are then determined by two parameters---the center-of-mass velocity $\mathbf{u} = \dot{\mathbf{R}}_{\mathrm{cm}}$ and the angular-velocity vector $\boldsymbol{\nu}$---and are given by
\begin{equation}
  \label{eqRigidMotionV}
  \dot{\mathbf{R}}_{\nucidxA}= \mathbf{u} + \boldsymbol{\nu} \times \Delta\mathbf{R}_{\nucidxA},
\end{equation}
where $\Delta\mathbf{R}_{\nucidxA}  = \mathbf{R}_{\nucidxA}  - \mathbf{R}_{\mathrm{cm}}$ are the nuclear coordinates relative to the center of mass.
For a uniform magnetic field, we obtain by summing over all nuclei in Eq.~\eqref{eqNewtonEOM} and using the expression in Eq.~\eqref{eqRigidMotionV} for the nuclear velocities, the equations
\begin{equation}
    \label{eqNewtonCMformOne}
  \begin{split}
\sum_{\nucidxA} \mass_{\nucidxA} \ddot{\mathbf{R}}_{\nucidxA}
 &= \sum_{\nucidxA} \charge_{\nucidxA} (\mathbf{u} + \boldsymbol{\nu}\times\Delta\mathbf{R}_{\nucidxA})\times\mathbf{B}
    \\
    & \ \ \ \ +  \sum_{\nucidxA\nucidxB} \BerryCurv_{\nucidxA\nucidxB} (\mathbf{u} + \boldsymbol{\nu}\times\Delta\mathbf{R}_{\nucidxB}),
  \end{split}
\end{equation}
assuming that the potential energy surface is translationally invariant,
$\sum_I \partial v(\mathbf{R})/\partial \mathbf R_I = \mathbf 0$. Next, introducing the total mass $m_{\mathrm{tot}}$, total charge $q_{\mathrm{tot}}$ and the center of nuclear charge $\mathbf{R}_{\mathrm{cc}} = (1/\charge_{\mathrm{tot}}) \sum_{\nucidxA} \charge_{\nucidxA} \mathbf{R}_{\nucidxA}$, we arrive at the following expression for the translational motion of the system
\begin{equation}
    \label{eqNewtonCMform}
  \begin{split}
    m_{\mathrm{tot}} \dot{\mathbf{u}} & = 
    q_{\mathrm{tot}} (\mathbf{u} + \boldsymbol{\nu} \times (\mathbf{R}_{\mathrm{cc}} - \mathbf{R}_{\mathrm{cm}}))\times\mathbf{B}
    \\
    & \ \ \ \ +  \sum_{\nucidxA\nucidxB} \BerryCurv_{\nucidxA\nucidxB} \mathbf{u}  + \sum_{\nucidxA\nucidxB} \BerryCurv_{\nucidxA\nucidxB} (\boldsymbol{\nu}\times\Delta\mathbf{R}_{\nucidxB}),
  \end{split}
\end{equation}
We note that the magnetic field and the Berry curvature couple the center-of-mass motion and the angular velocity of the system. No such coupling occurs in the absence of a magnetic field, if (as is typical) the wave function can be chosen to be real valued.

To obtain the corresponding equation for the angular velocity of the system, we evaluate the torque on each nucleus by taking the cross product of Eq.~\eqref{eqNewtonEOM} with $\Delta\mathbf{R}_{\nucidxA}$:
\begin{equation}
\label{eqTorqueRI}
  \begin{split}
     \Delta\mathbf{R}_{\nucidxA}&\times \mass_{\nucidxA} \ddot{\mathbf{R}}_{\nucidxA} \\
&= - \Delta\mathbf{R}_{\nucidxA} \times \frac{\partial\PES}{\partial \mathbf{R}_{\nucidxA}} \\&\quad + \charge_{\nucidxA} \Delta\mathbf{R}_{\nucidxA} \times ((\mathbf{u} + \boldsymbol{\nu}\times \Delta\mathbf{R}_{\nucidxA})\times\mathbf{B})
  \\
  & \qquad+ \sum_{\nucidxB} \Delta\mathbf{R}_{\nucidxA}\times \BerryCurv_{\nucidxA\nucidxB} (\mathbf{u} + \boldsymbol{\nu}\times\Delta\mathbf{R}_{\nucidxB}).
  \end{split}
\end{equation}
We next express the acceleration $\ddot{\mathbf{R}}_{\nucidxA}$ in terms of the angular velocity by taking the time derivative of $\dot{\mathbf{R}}_{\nucidxA}$ in \eqref{eqRigidMotionV}, yielding:
\begin{equation}
  \label{eqRigidMotionA}
  \ddot{\mathbf{R}}_{\nucidxA} = \dot{\mathbf{u}} + \dot{\boldsymbol{\nu}} \times \Delta\mathbf{R}_{\nucidxA} + \boldsymbol{\nu} \times (\boldsymbol{\nu} \times \Delta\mathbf{R}_{\nucidxA}).
\end{equation}
Substituting this result into Eq.~\eqref{eqTorqueRI},
summing over all nuclei, introducing the moment-of-inertia tensor $\Inertia$ and the electrical quadrupole moment tensor $\QuadMom$ given by
\begin{align}
\label{eqInertMom}
\Inertia &= \sum_{\nucidxA} \mass_{\nucidxA} (|\Delta\mathbf{R}_{\nucidxA}|^2 \mathbf{I} - \Delta\mathbf{R}_{\nucidxA} \Delta\mathbf{R}_{\nucidxA}^\mathrm T), \\
\QuadMom &= \sum_{\nucidxA} \charge_{\nucidxA} (|\Delta\mathbf{R}_{\nucidxA}|^2 \mathbf{I} - \Delta\mathbf{R}_{\nucidxA} \Delta\mathbf{R}_{\nucidxA}^\mathrm T),
\end{align}
and assuming that the potential energy surface gives rise to no net torque, 
$\sum_I m_I \Delta\mathbf{R}_{\nucidxA} \times {\partial\PES}/{\partial \mathbf{R}_{\nucidxA}} = \mathbf 0$,
we obtain:
\begin{equation}
  \label{eqEulerEOM}
  \begin{split}
    \Inertia \dot{\boldsymbol{\nu}} & + (\Inertia \boldsymbol{\nu}) \times \boldsymbol{\nu}
    \\
  & = \charge_{\mathrm{tot}} (\mathbf{R}_{\mathrm{cc}}-\mathbf{R}_{\mathrm{cm}})\times(\mathbf{u}\times\mathbf{B}) - (\QuadMom \mathbf{B}) \times \boldsymbol{\nu}
  \\
  & \ \ \ + \sum_{\nucidxA\nucidxB} \Delta\mathbf{R}_{\nucidxA}\times \BerryCurv_{\nucidxA\nucidxB} (\mathbf{u} + \boldsymbol{\nu}\times\Delta\mathbf{R}_{\nucidxB}).
  \end{split}
\end{equation}
This equation determines the angular acceleration $\dot{\boldsymbol{\nu}}$ or, if one prefers, the (closely related) time derivative of the angular momentum. It is a form of Euler's rigid body equation of motion, supplemented with the effects of the Berry screening force.

The coupling between the translational and rotational  equations of motion in Eqs.~\eqref{eqNewtonCMform} and~\eqref{eqEulerEOM}, respectively, becomes more transparent by mapping one of the vectors in a cross product to a skew symmetric matrix according to 
\begin{equation}
\mathbf{s}\times\mathbf{t} = \SkewSymEncode{\mathbf{s}} \mathbf{t} = -\SkewSymEncode{\mathbf{t}} \mathbf{s},
\end{equation}
in the notation
\begin{equation}
  \SkewSymEncode{\mathbf{s}} = \begin{pmatrix} 0 & -s_3 & s_2 \\ s_3 & 0 & -s_1 \\ -s_2 & s_1 & 0 \end{pmatrix}
  , \quad \mathbf{s} = (s_1,s_2,s_3).
\end{equation}
Rewriting Eq.~\eqref{eqNewtonCMform} and \eqref{eqEulerEOM} and introducing the $3\times 3$ translational--rotational coupling matrix
\begin{equation}
\label{eqGmat}
 \boldsymbol { \mathcal{G} }= q_{\mathrm{tot}} \SkewSymEncode{\mathbf{B}} \SkewSymEncode{\mathbf{R}_{\mathrm{cc}} - \mathbf{R}_{\mathrm{cm}}}  - \sum_{\nucidxA\nucidxB} \BerryCurv_{\nucidxA\nucidxB} \SkewSymEncode{\Delta\mathbf{R}_{\nucidxB}},
\end{equation}
Newton's equation for center-of-mass motion becomes
\begin{equation}
    \label{eqNewtonCMalt}
    m_{\mathrm{tot}} \dot{\mathbf{u}}  =  -q_{\mathrm{tot}} \mathbf B \times \mathbf u + \sum_{\nucidxA\nucidxB} \BerryCurv_{\nucidxA\nucidxB} \, \mathbf{u}+ \boldsymbol{ \mathcal{G}} \boldsymbol{\nu},
\end{equation}
while Euler's equation for rotation about the center-of-mass becomes
\begin{equation}
  \label{eqEulerEOMalt}
  \begin{split}
    \!\Inertia \dot{\boldsymbol{\nu}} + (\Inertia \boldsymbol{\nu}) \times \boldsymbol{\nu} &= -(\QuadMom \mathbf{B}) \times \boldsymbol{\nu}
    \\
  &  - \sum_{\nucidxA\nucidxB} \SkewSymEncode{\Delta\mathbf{R}_{\nucidxA}} \BerryCurv_{\nucidxA\nucidxB} \SkewSymEncode{\Delta\mathbf{R}_{\nucidxB}} \boldsymbol{\nu}-\boldsymbol{\mathcal{G}}^\mathrm T \mathbf{u}.
  \end{split}
\end{equation}

From Eqs.\,\eqref{eqNewtonCMform} and \eqref{eqNewtonCMalt}, it is  clear that the center of mass acceleration is affected by the internal rotation $\boldsymbol{\nu}$. The strength of this coupling is proportional to the difference between the center of charge and the center of mass as well as to the magnetic field, although it may be partially counterbalanced by the Berry curvature term. Likewise, from Eqs.\,\eqref{eqEulerEOM} and \eqref{eqEulerEOMalt}, it is clear that the angular acceleration of the internal rotation is affected by the center of mass motion. Again the coupling is proportional to the the difference between the center of charge and the center of mass as well as to the magnetic field, and it may be counterbalanced by the effect of the Berry curvature.

\subsection{Equation of motion for small oscillations}

Next, we choose the initial position to be a local minimum $\mathbf{R}_0$ on the potential energy surface $\PES(\mathbf{R})$, so that $\partial\PES/\partial\mathbf{R}=\mathbf{0}$. We assume that all time dependence is captured by linear motion and an oscillatory displacement $\epsilon \DisplVec(t)$ that is small enough for our below approximations to be valid:
\begin{equation}
\label{eqSmallOsc}
\mathbf{R}(t) = \mathbf{R}_0 + \dot{\mathbf{R}}_0 t + \epsilon \DisplVec(t).
\end{equation}
Denoting the Hessian and the term arising from geometrical gradients of the Berry curvature tensor at $\mathbf{R}=\mathbf{R}_0$ by
\begin{align}
  \HessianComp_{\nucidxA\cartidxA,\nucidxB\cartidxB} & = \left.\frac{\partial^2 \PES(\mathbf{R})}{\partial R_{\nucidxA\cartidxA} \partial R_{\nucidxB\cartidxB}}\right\vert_{\mathbf R = \mathbf R_0},
  \\
  \TotRotGradComp_{\nucidxA\cartidxA,\nucidxB\cartidxB} & = \sum_{\nucidxC\cartidxC} \left.\frac{\partial\TotRotComp_{\nucidxA\cartidxA,\nucidxC\cartidxC}(\mathbf{R})}{\partial R_{\nucidxB\cartidxB}}\right\vert_{\mathbf R = \mathbf R_0} \!\!\!\dot{R}_{0;\nucidxC\cartidxC}\,,
   \label{eqLambdaAsTotRotGrad}
\end{align}
and letting $\TotRot = \TotRot(\mathbf{R}_0)$,
we may write the equations of motion in Eq.~\eqref{eqmotion} to
first order in $\epsilon$ in the manner
\begin{equation}
  \label{eqOscTEOM}
  \mathbf{M} \ddot{\DisplVec} = (\TotRotGrad-\Hessian) \DisplVec + \TotRot \dot{\DisplVec}.
\end{equation}
Remarkably, the gradients of
$\TotRot(\mathbf{R})$ couple with the linear motion
$\dot{\mathbf{R}}_0$ to produce a correction to the Hessian. In what
follows, we assume that this correction vanishes,
either because $\TotRot(\mathbf{R})$ is approximately constant around
$\mathbf{R} = \mathbf{R}_0 + \dot{\mathbf{R}}_0 t + \epsilon \DisplVec$ or because there is no linear motion, $\dot{\mathbf{R}}_0 = \mathbf{0}$.

A Fourier transformation of Eq.~\eqref{eqOscTEOM} with $\boldsymbol \Lambda = \mathbf 0$ yields 
\begin{equation}
  \label{eqQEP}
 (\Hessian - \mathrm i\omega\TotRot -\omega^2 \mathbf{M}) \hat{\DisplVec}(\omega) = \mathbf{0}
\end{equation}
or, equivalently,
\begin{equation}
\label{eqMWQEP}  
 (\mwHessian - \mathrm i\omega \mwTotRot -\omega^2 \mathbf{I}) \, \hat { \boldsymbol{\zeta}}(\omega) = \mathbf{0},
\end{equation}
in terms of the mass-weighted Hessian and  mass-weighted Berry curvature
\begin{align}
\mwHessian &= \mathbf{M}^{-1/2} \Hessian \mathbf{M}^{-1/2}, \\
\mwTotRot &= \mathbf{M}^{-1/2} \TotRot \mathbf{M}^{-1/2},
\end{align}
and $\hat{ \boldsymbol{\zeta}}(\omega) = \mathbf{M}^{1/2} \hat \DisplVec(\omega)$.
Because $\omega$ enters both linearly and quadratically, these equations are instances of the quadratic eigenvalue problem (QEP)~\cite{TISSEUR_SIAMR43_235} rather than of a standard eigenvalue problem. In principle, even the usual vibrational eigenvalue problem without velocity-dependent forces, $\Hessian \hat{\DisplVec} = \omega^2 \mathbf{M} \hat{\DisplVec}$, is a QEP since $\omega$ enters quadratically. However, it is trivially re-expressed as a standard eigenvalue problem of the same dimension by introducing $\lambda = \omega^2$.

A general QEP can be transformed to an equivalent standard eigenvalue problem of doubled dimension. Such a transformation is said to \emph{linearize} the QEP. There are many ways of linearizing a QEP and here we illustrate one of them. By introducing a new variable for the velocity, $\DisplDeriv(t) = \dot{\DisplVec}(t)$, the equation of motion can be written in a form that is first-order in time:
\begin{align}
  \DisplDeriv & = \dot{\DisplVec}, \\
  \mathbf{M} \dot{\DisplDeriv} & = -\Hessian \DisplVec + \TotRot \DisplDeriv.
\end{align}
In the frequency domain, the corresponding system is
\begin{align}
  \hat{\DisplDeriv}(\omega) & = \mathrm i\omega \hat{\DisplVec}(\omega), \\
\mathrm  i\omega \mathbf{M} \hat{\DisplDeriv}(\omega) & = -\Hessian \hat{\DisplVec}(\omega) + \TotRot \hat{\DisplDeriv}(\omega).
\end{align}
Finally, these equations can be rearranged into a generalized eigenvalue problem
\begin{equation}
  \label{eqSEP}
  \begin{pmatrix}
      \mathbf{0} & \mathbf{I} \\
      -\Hessian & \TotRot
  \end{pmatrix}
  \begin{pmatrix}
      \hat{\DisplVec} \\ \hat{\boldsymbol{\gamma}}
  \end{pmatrix}
  =\mathrm  i\omega
  \begin{pmatrix}
      \mathbf{I} & \mathbf{0} \\
      \mathbf{0} & \mathbf{M}
  \end{pmatrix}
  \begin{pmatrix}
      \hat{\DisplVec} \\ \hat{\boldsymbol{\gamma}}
  \end{pmatrix}.
\end{equation}
Alternatively, in terms of the mass-weighted quantities $\boldsymbol{\zeta} = \mathbf{M}^{1/2} \DisplVec$ and $\boldsymbol{\xi} = \dot{\boldsymbol{\zeta}} = \mathbf{M}^{1/2} \dot{\DisplVec}$, we obtain the standard eigenvalue problem
\begin{equation}
  \label{eqMWSEP}
  \begin{pmatrix}
      \mathbf{0} & \mathbf{I} \\
      -\mwHessian & \mwTotRot
  \end{pmatrix}
  \begin{pmatrix}
      \hat{\boldsymbol{\zeta}} \\ \hat{\boldsymbol{\xi}}
  \end{pmatrix}
  = \mathrm i\omega
  \begin{pmatrix}
      \hat{\boldsymbol{\zeta}} \\ \hat{\boldsymbol{\xi}}
  \end{pmatrix}.
\end{equation}
Compared to a standard eigenvalue problem of the same dimension, a quadratic eigenvalue problem has twice as many eigenvalues, although two distinct eigenvalues may share the same eigenvector. Solving one of the quadratic eigenvalue problems in Eq.~\eqref{eqQEP}--\eqref{eqMWQEP} of dimension $3N$ or one of the standard eigenvalue problems in Eq.~\eqref{eqSEP}-\eqref{eqMWSEP} of doubled dimension $6N$ yields a discrete set of $6N$ dynamical frequencies $\omega_k$ and oscillatory modes $\hat{\DisplVec}_k = \mathbf{M}^{-1/2} \hat{\boldsymbol{\zeta}}_k$.

Returning to the original QEP, we note three characterizations of the eigenmodes. Firstly, because the matrices $\Hessian,\TotRot,\mathbf{M}$ are all real, taking the complex conjugate of Eq.~\eqref{eqQEP} leads to the conclusion that the eigenmodes come in complex conjugated pairs: \textit{If $\hat{\DisplVec}_k$ is a solution with frequency $\omega_k$, then $\hat{\DisplVec}_k^*$ is also a solution with frequency $-\omega_k^*$.} For simplicity we restrict attention to real frequencies from now on and thus expect $\hat{\DisplVec}_k(\omega_k) = \hat{\DisplVec}_k(-\omega_k)^*$ to hold.

Secondly, if also the eigenmode $\hat{\DisplVec}_k$ is real, then it satisfies the two equations
  \begin{align}
    (\Hessian - \omega_k^2 \mathbf{M}) \hat{\DisplVec}_k& = 0,
    \\
      \omega_k \TotRot \hat{\DisplVec}_k & = 0,
  \end{align}
  separately. \textit{Hence, real modes must either have zero frequency $\omega_k=0$ or belong to the null space of $\TotRot$. Moreover, real modes must also be solutions of a vibrational problem without velocity dependent forces.}

Thirdly, some modes have two distinct frequencies. Equivalently, suppose the modes are ordered so that $\omega_1$ and $\omega_2$ share the same mode $\hat{\DisplVec} = \hat{\DisplVec}_1 = \hat{\DisplVec}_2$. This mode satisfies
\begin{align}
  (\Hessian - \mathrm i \omega_1 \TotRot - \omega_1^2 \mathbf{M}) \hat{\DisplVec} & = 0,
  \\
  (\Hessian - \mathrm i \omega_2 \TotRot - \omega_2^2 \mathbf{M}) \hat{\DisplVec} & = 0,
\end{align}
simultaneously. Subtracting one equation from the other and dividing by $\omega_2-\omega_1\neq 0$ yields
\begin{align}
  \label{eqQEPTwoFreqModeCond}
  ( \mathrm i\TotRot + (\omega_1 + \omega_2) \mathbf{M}) \hat{\DisplVec} & = 0.
\end{align}
\textit{Hence, eigenmodes with more than one frequency must be generalized eigenvectors of $\TotRot$. Moreover, if the corresponding eigenvalue is different from zero (i.e., if $\omega_1+\omega_2 \neq 0$), then the eigenmode cannot be real valued.}

Regarding distinct frequencies that share the same eigenvector, we stress  that it is the pair of a frequency and an eigenvector that together determine the dynamics in the time domain---whereas the pair $(\omega_1, \hat{\DisplVec})$ corresponds to $\DisplVec_1(t) = 2\mathrm{Re}(\hat{\DisplVec} \mathrm e^{\mathrm i\omega_1 t})$, the pair $(\omega_2, \hat{\DisplVec})$ corresponds to $\DisplVec_2(t) = 2\mathrm{Re}(\hat{\DisplVec} \mathrm e^{\mathrm i\omega_2 t})$.

\subsection{A frequency-dependent orthogonality property}
\label{secQEPFreqOrtho}

The standard vibrational problem without velocity-dependent forces yields $3N$ normal modes that are orthogonal with the mass tensor $\mathbf{M}$ as the metric. The QEP yields $6N$ modes $\hat{\DisplVec}_k$ and it is immediately clear that they cannot all be orthogonal. Projecting Eq.~\eqref{eqQEP} from the right by a different mode yields
\begin{equation}
  \hat{\DisplVec}_l^{\dagger} (\Hessian - \mathrm{i} \omega_k \TotRot - \omega_k^2 \mathbf{M}) \hat{\DisplVec}_k = 0
\end{equation}
and taking the complex conjugate and using that $\TotRot^T = -\TotRot$ yields
\begin{equation}
  \hat{\DisplVec}_k^{\dagger} (\Hessian - \mathrm{i} \omega_k \TotRot - \omega_k^2 \mathbf{M}) \hat{\DisplVec}_l = 0.
\end{equation}
Setting $k=1,l=2$ in the first equation and $k=2,l=1$ in the second equation and taking the difference now yields
\begin{equation}
  \hat{\DisplVec}_2^{\dagger} \left((\omega_2^2 - \omega_1^2) \mathbf{M} + \mathrm{i} (\omega_2-\omega_1) \TotRot\right) \hat{\DisplVec}_1 = 0.
\end{equation}
When $\omega_1 \neq \omega_2$ this simplifies to
\begin{equation}
  \hat{\DisplVec}_2^{\dagger} \left((\omega_2 + \omega_1) \mathbf{M} + \mathrm{i} \TotRot\right) \hat{\DisplVec}_1 = 0.
\end{equation}
For real modes we may discard the $\TotRot$ term and recover the usual orthogonality property. However, complex modes are in general not orthogonal in the metric $\mathbf{M}$ nor in any other fixed metric.

\subsection{Configuration-space rotations and rovibrational coupling}

In the time domain, Eq.~\eqref{eqQEPTwoFreqModeCond} corresponds to
\begin{equation}
  \mathbf{M} \ddot{\DisplVec} = \TotRot \dot{\DisplVec}.
\end{equation}
The same simplified equation of motion also arises in the special case when the Hessian can be neglected, $\Hessian \approx \mathbf{0}$. This equation of motion can be solved explicitly since the real antisymmetric matrix $\TotRot$ is a generator of rotations in the $3N$-dimensional configuration space,
\begin{align}
  \DisplVec(t) & = \mathbf{M}^{-1/2}  \, \mathrm e^{\mwTotRot t} \, \mathbf{M}^{1/2} \DisplVec(0),
  \\
  \mathbf{R}(t) & = \mathbf{R}_0+ \epsilon \mathbf{M}^{-1/2}  \, \mathrm e^{\mwTotRot t} \, \mathbf{M}^{1/2} \DisplVec(0),
\end{align}
with $\mwTotRot = \mathbf{M}^{-1/2} \TotRot \mathbf{M}^{-1/2}$.
In general, rotations in configuration space preserve all distances and angles in configuration space, but do not necessarily correspond to rigid rotations that preserve internuclear distances $|\mathbf{R}_{\nucidxA} - \mathbf{R}_{\nucidxB}|$ in three-dimensional space. Hence, even for a vanishing Hessian, rovibrational coupling arises due to $\TotRot$.

An instructive special case arises when the Berry curvature vanishes, $\BerryCurv = \mathbf{0}$, so that the external Lorentz force is the only velocity-dependent force. Then $\smash{\mwTotRot_{\nucidxA\nucidxB} = ({\charge_{\nucidxA}}/{\mass_{\nucidxA}}) \delta_{\nucidxA\nucidxB} \boldsymbol{\nu}}$ has a block-diagonal structure with  $\nu_{\cartidxA\cartidxB} = \epsilon_{\cartidxA\cartidxB\cartidxC} B_{\cartidxC}$. The motion of each nucleus can now be solved independently,
\begin{align}
  \DisplVec_{\nucidxA}(t) & = \mathrm e^{\frac{\charge_{\nucidxA}}{\mass_{\nucidxA}} \boldsymbol{\nu} t} \, \DisplVec_{\nucidxA}(0),
  \\
  \mathbf{R}_{\nucidxA}(t) & = \mathbf{R}_{0;\nucidxA} + \epsilon\, \mathrm e^{\frac{\charge_{\nucidxA}}{\mass_{\nucidxA}} \boldsymbol{\nu} t}\, \DisplVec_{\nucidxA}(0),
\end{align}
As a result, each nucleus undergoes independent rotation with an angular velocity that depends on its gyromagnetic ratio $\charge_{\nucidxA}/\mass_{\nucidxA}$. For this to be a rigid rotation that preserves all pairwise distances $|\mathbf{R}_I(t) - \mathbf{R}_J(t)|$, the gyromagnetic ratios must be identical, $\charge_{\nucidxA}/\mass_{\nucidxA} = \charge_{\nucidxB}/\mass_{\nucidxB}$. Additionally, the nuclei must either share the same gyrocenter, $\mathbf{R}_{0;\nucidxA} = \mathbf{R}_{0;\nucidxB}$, or all motions must be synchronized in the sense that the initial displacements are identical, $\DisplVec_{\nucidxA}(0) = \DisplVec_{\nucidxB}(0)$. In other cases, configuration-space rotations couple rigid rotations and vibrational motion.

\subsection{Quantification of translational, rotational, and vibrational coupling}

A mode $\hat{\DisplVec}(\omega_k)$ can be decomposed into components that correspond to rigid translation, rigid rotation, and internal motion, respectively. Here we are interested in the kinetic energies associated with these components. From Eq.\,\eqref{eqSmallOsc},
we obtain for the $k$th mode the following Cartesian displacement of nucleus $I$:
\begin{equation}
  \begin{split}
    \mathbf{R}_{\nucidxA}(t) & = \mathbf{R}_{0;\nucidxA} + \epsilon \DisplVec_{k;\nucidxA}(t)
  \end{split}
\end{equation}
where 
\begin{equation}
\DisplVec_{k;\nucidxA}(t) = \hat{\DisplVec}_{k;\nucidxA} \mathrm e^{\mathrm i\omega_k t} + \hat{\DisplVec}_{k;\nucidxA}^* \mathrm e^{-\mathrm i\omega_k t} 
\end{equation}
The kinetic energy of this mode is given by
\begin{equation}
  \begin{split}
    T_k & = \frac{\epsilon^2}{2} \sum_{\nucidxA} \mass_{\nucidxA} |\dot{\DisplVec}_{k;\nucidxA}|^2
    \\
    &  = \frac{\epsilon^2 \omega_k^2}{2} \sum_{\nucidxA} \mass_{\nucidxA} \big(2\hat{\DisplVec}_{k;\nucidxA}^*\cdot \hat{\DisplVec}_{k;\nucidxA} - \hat{\DisplVec}_{k;\nucidxA}\cdot \hat{\DisplVec}_{k;\nucidxA} \mathrm e^{2\mathrm i\omega_k t}
    \\
    & \quad\qquad \qquad\qquad \qquad \qquad - \hat{\DisplVec}_{k;\nucidxA}^* \cdot \hat{\DisplVec}_{k;\nucidxA}^* \mathrm e^{-2\mathrm i\omega_k t} \big).
  \end{split}
\end{equation}
The kinetic energy is thus a periodic function of time, with time average
\begin{equation}
  \begin{split}
    \overline {T}_k & = \epsilon^2 \omega_k^2 \sum_{\nucidxA} \mass_{\nucidxA} |\hat{\DisplVec}_{k;\nucidxA}|^2.
  \end{split}
\end{equation}
Similarly, the total mechanical momentum is
\begin{equation}
  \begin{split}
    \overline{\boldsymbol{\Pi}} & = \epsilon \sum_{\nucidxA} \mass_{\nucidxA} \dot{\DisplVec}_{k;\nucidxA} \\ &= \mathrm i\epsilon \omega_k \sum_{\nucidxA} \mass_{\nucidxA} (\hat{\DisplVec}_{k;\nucidxA} \mathrm e^{\mathrm i\omega_k t} - \hat{\DisplVec}_{k;\nucidxA}^* \mathrm e^{-\mathrm i\omega_k t})
  \end{split}
\end{equation}
and the time average of the kinetic energy due to the center-of-mass motion is therefore
\begin{equation}
  \begin{split}
    \overline{T}_{\text{cm}} & = \frac{\epsilon^2 \omega_k^2}{\TotMass} \Bigl| \sum_{\nucidxA} \mass_{\nucidxA} \hat{\DisplVec}_{k;\nucidxA} \Bigr|^2.
  \end{split}
\end{equation}
Next, we consider the total mechanical angular momentum $\mathbf{J}$ relative to the center of mass $\mathbf{R}_{\text{cm}}$ of the equilibrium geometry. With the notation $\Delta \mathbf{R}_{0;\nucidxA} = \mathbf{R}_{0;\nucidxA} - \mathbf{R}_{\text{cm}}$, we may write
\begin{equation}
  \begin{split}
    \mathbf{J} & = \sum_{\nucidxA} (\Delta \mathbf{R}_{0;\nucidxA}+ \epsilon \DisplVec_{k;\nucidxA}) \times \mass_{\nucidxA} \dot{\DisplVec}_{k;\nucidxA}
    \\
    & = \sum_{\nucidxA} \left(\Delta \mathbf{R}_{0;\nucidxA}+ \epsilon (\hat{\DisplVec}_{k;\nucidxA} \mathrm e^{\mathrm i\omega_k t}+\hat{\DisplVec}_{k;\nucidxA}^* \mathrm e^{-\mathrm i\omega_k t})\right)
    \\
    & \quad \qquad \quad \times \mass_{\nucidxA} \, \mathrm i\epsilon \omega_k (\hat{\DisplVec}_{k;\nucidxA} \mathrm e^{\mathrm i\omega_k t}-\hat{\DisplVec}_{k;\nucidxA}^* \mathrm e^{-\mathrm i\omega_k t}).
  \end{split}
\end{equation}
Introducing the auxiliary quantity
\begin{equation}
  \hat{\mathbf{K}}  = \sum_{\nucidxA} \mass_{\nucidxA} \Delta \mathbf{R}_{0;\nucidxA} \times \hat{\DisplVec}_{k;\nucidxA},
\end{equation}
the time average of the rotational energy becomes, to second order in $\epsilon$, 
\begin{equation}
  \begin{split}
    \overline{T}_{\text{rot}} & =  \epsilon^2 \omega_k^2 \hat{\mathbf{K}}^{\dagger} \Inertia^{-1} \hat{\mathbf{K}},
  \end{split}
\end{equation}
where $\Inertia$ is the moment-of-inertia tensor of Eq.\,\eqref{eqInertMom}.

As a simple measure of the degree to which a mode represents center-of-mass motion, rotation about the center of mass, and vibration, we define the fractions
\begin{align}
  P_{\text{cm}}& = \frac{\overline{T}_{\text{cm}}}{\overline{T}}, \\
  P_{\text{rot}} & = \frac{\overline{T}_{\text{rot}}}{\overline{T}}, \\
  P_{\text{vib}} & = \frac{\overline{T}-\overline{T}_{\text{cm}}-\overline{T}_{\text{rot}}}{\overline{T}} = 1 - P_{\text{cm}} - P_{\text{rot}},
\end{align}
which are nonnegative and add up to 1. Numerical examples will be
given below, in Sec.~\ref{secResults}. Whereas the interpretation of
$P_{\text{rot}}$ is relatively straightforward, it should be kept in
mind that $P_{\text{cm}}$ measures both linear translation and
cyclotron-like motion of the center of mass. Both $P_{\text{rot}}$ and
$P_{\text{cm}}$ capture only rigid motion and the non-rigid motion is
quantified by $P_{\text{vib}}$.

\section{Hamiltonian formulation}
\label{secHamiltonian}

It is straightforward to generalize the above treatment based on Newton's equation of motion to a Lagrangian or Hamiltonian formulation more suitable to non-Cartesian coordinates. Here, however, we highlight different aspects of in particular the Hamiltonian formulation: firstly, its lack of gauge invariance and, secondly, that it leads directly to a linearized form of the QEP.

For convenience, we let $\mathbf{a}_{\nucidxA}(\mathbf{R}) = -q_{\nucidxA} \mathbf{A}(\mathbf{R}_{\nucidxA}) + \GVP_{\nucidxA}(\mathbf{R})$ denote a three-dimensional vector as function of the configuration $\mathbf{R}$ and $\mathbf{a}(\mathbf{R})$ denote the corresponding $3N$-dimensional vector. In Cartesian coordinates $\mathbf{R}$, the nuclear Hamiltonian may then be written
\begin{equation}
  \mathcal{H}(\mathbf{P},\mathbf{R}) = \frac{1}{2} (\mathbf{P} + \mathbf{a}(\mathbf{R}))^\mathrm T \mathbf{M}^{-1} (\mathbf{P} + \mathbf{a}(\mathbf{R})) + v(\mathbf{R}).
\end{equation}
The Hamiltonian is already quadratic in the momenta $\mathbf{P}$. To construct an approximation that is second order in the pair $(\mathbf{R},\mathbf{P})$, we consider the Taylor expansion around a point $\mathbf{R}_0$ of vanishing gradient $\partial v/\partial \mathbf{R}$. Defining
\begin{align}
  F_{\nucidxB\cartidxB, \nucidxA\cartidxA}(\mathbf{R}_0) & = \left. \frac{\partial a_{\nucidxB\cartidxB}(\mathbf{R})}{\partial R_{\nucidxA\cartidxA}} \right|_{\mathbf{R}=\mathbf{R}_0},
  \\
  S_{\nucidxA\cartidxA, \nucidxC\cartidxC}(\mathbf{R}_0) & = \sum_{\nucidxB\cartidxB} a_{\nucidxB\cartidxB}(\mathbf{R}) \mass_{\nucidxB}^{-1} \left. \frac{\partial^2 a_{\nucidxB\cartidxB}(\mathbf{R})}{\partial R_{\nucidxA\cartidxA} \partial R_{\nucidxC\cartidxC}} \right|_{\mathbf{R}=\mathbf{R}_0},
   \label{eqSDEF}
\end{align}
we have
\begin{align}
  v(\mathbf{R}_0+\mathbf{s}) & \approx v(\mathbf{R}_0) + \frac{1}{2} \mathbf{s}^\mathrm T \Hessian \mathbf{s},
  \\
  \mathbf{a}(\mathbf{R}_0+\mathbf{s}) & \approx a(\mathbf{R}_0) + \mathbf{F}(\mathbf{R}_0) \, \mathbf{s},
\end{align}
and
\begin{equation}
  \begin{split}
   & \mathbf{a}(\mathbf{R}_0+\mathbf{s})^\mathrm T \mathbf{M}^{-1} \mathbf{a}(\mathbf{R}_0+\mathbf{s}) \approx \mathbf{a}(\mathbf{R}_0)^\mathrm T \mathbf{M}^{-1} \mathbf{a}(\mathbf{R}_0)
    \\
    & \qquad\qquad +  2 \mathbf{a}(\mathbf{R}_0)^\mathrm T \mathbf{M}^{-1} \mathbf{F}(\mathbf{R}_0) \, \mathbf{s} + \frac{1}{2} \mathbf{s}^\mathrm T \mathbf{S}(\mathbf{R}_0) \mathbf{s}.
  \end{split}
\end{equation}
In terms of the $6N\times 6N$ matrix
\begin{equation}
  \begin{split}
    \mathcal{H}^{[2]}_{\text{mat}} (\mathbf{R}_0) & = 
    \begin{pmatrix}
      \Hessian + \mathbf{S} + \mathbf{F}^\mathrm T \mathbf{M}^{-1} \mathbf{F}  & \mathbf{F}^\mathrm T \mathbf{M}^{-1} \\
      \mathbf{M}^{-1} \mathbf{F} & \mathbf{M}^{-1}
    \end{pmatrix}
  \end{split}
\end{equation}
the Hamiltonian can now be written as
\begin{equation}
  \mathcal{H}(\mathbf{P},\mathbf{R}) \approx \frac{1}{2}
  \begin{pmatrix} \mathbf{R}-\mathbf{R}_0 \\ \mathbf{P} + \mathbf{a}(\mathbf{R}_0) \end{pmatrix}^\mathrm T
  \mathcal{H}^{[2]}_{\text{mat}}
  \begin{pmatrix} \mathbf{R}-\mathbf{R}_0 \\ \mathbf{P} + \mathbf{a}(\mathbf{R}_0) \end{pmatrix}
\end{equation}
and Hamilton's equations of motion take the form
\begin{equation}
  \label{eqHamiltonsEOM}
  \begin{split}
  \begin{pmatrix}
    \dot{\mathbf{R}} \\
    \dot{\mathbf{P}}
  \end{pmatrix}
  & =
  \begin{pmatrix} \mathbf{0} & \mathbf{I} \\ -\mathbf{I} & \mathbf{0} \end{pmatrix}
  \begin{pmatrix}
    \partial\mathcal{H} / \partial \mathbf{R} \\
    \partial\mathcal{H} / \partial \mathbf{P}    
  \end{pmatrix}
  \\
  & = \begin{pmatrix} \mathbf{0} & \mathbf{I} \\ -\mathbf{I} & \mathbf{0} \end{pmatrix} \mathcal{H}^{[2]}_{\text{mat}}\begin{pmatrix} \mathbf{R}-\mathbf{R}_0 \\ \mathbf{P} + \mathbf{a}(\mathbf{R}_0) \end{pmatrix}
  \end{split}
\end{equation}
The ansatz $\mathbf{R}(t) = \mathbf{R}_0 + \epsilon \DisplVec(t)$ and $\mathbf{P}(t) = -\mathbf{a}(\mathbf{R}_0) + \epsilon \boldsymbol{\kappa}(t)$ now yields, to first order in $\epsilon$, the equations of motion
\begin{equation}
  \begin{pmatrix}
    \dot{\DisplVec} \\ \dot{\boldsymbol{\kappa}}
  \end{pmatrix}
   = \begin{pmatrix} \mathbf{0} & \mathbf{I} \\ -\mathbf{I} & \mathbf{0} \end{pmatrix}  \mathcal{H}^{[2]}_{\text{mat}}
  \begin{pmatrix}
    \DisplVec \\ \boldsymbol{\kappa}
  \end{pmatrix}.
\end{equation}
Finally, Fourier transformation yields the eigenvalue equation
\begin{subequations}
\begin{equation}
  \label{eqHamLinearized}
  \mathrm i\omega
  \begin{pmatrix}
    \hat{\DisplVec}(\omega) \\ \hat{\boldsymbol{\kappa}}(\omega)
  \end{pmatrix}
   = \begin{pmatrix} \mathbf{0} & \mathbf{I} \\ -\mathbf{I} & \mathbf{0} \end{pmatrix} \mathcal{H}^{[2]}_{\text{mat}}
  \begin{pmatrix}
    \hat{\DisplVec}(\omega) \\ \hat{\boldsymbol{\kappa}}(\omega)
  \end{pmatrix}
\end{equation}
or, equivalently,
\begin{equation}
  \label{eqHamLinearizedJasMetric}
 - \mathrm i\omega \begin{pmatrix} \mathbf{0} & \mathbf{I} \\ -\mathbf{I} & \mathbf{0} \end{pmatrix}
  \begin{pmatrix}
    \hat{\DisplVec}(\omega) \\ \hat{\boldsymbol{\kappa}}(\omega)
  \end{pmatrix}
   = \mathcal{H}^{[2]}_{\text{mat}}
  \begin{pmatrix}
    \hat{\DisplVec}(\omega) \\ \hat{\boldsymbol{\kappa}}(\omega)
  \end{pmatrix}.
\end{equation}
A change of variables first to mass-weighted coordinates and then to $\hat{\boldsymbol{\alpha}}_{\pm}(\omega) = \mathbf{M}^{1/2} \hat{\DisplVec}(\omega) \pm \mathrm{i} \mathbf{M}^{-1/2} \hat{\boldsymbol{\kappa}}(\omega)$ produces a third equivalent form,
\begin{equation}
  \omega \begin{pmatrix} \mathbf{I} & \mathbf{0} \\ \mathbf{0} & -\mathbf{I} \end{pmatrix} \begin{pmatrix} \hat{\boldsymbol{\alpha}}_{-}(\omega) \\ \hat{\boldsymbol{\alpha}}_{+}(\omega) \end{pmatrix}
  =
  \begin{pmatrix} \mathcal{A} & \mathcal{B} \\ \mathcal{B}^* & \mathcal{A}^* \end{pmatrix}
  \begin{pmatrix} \hat{\boldsymbol{\alpha}}_{-}(\omega) \\ \hat{\boldsymbol{\alpha}}_{+}(\omega) \end{pmatrix},
\end{equation}
\end{subequations}
where 
\begin{align}
  \mathcal{A} & = \mathbf{I} + \mathbf{M}^{-1/2}(\Hessian + \mathbf{S} + \mathbf{F}^T \mathbf{M}^{-1} \mathbf{F}) \mathbf{M}^{-1/2}
               \nonumber\\
             & \ \ \ -\mathrm{i} \mathbf{M}^{-1/2}  (\mathbf{F} - \mathbf{F}^T)\mathbf{M}^{-1/2},
  \\
  \mathcal{B} & = -\mathbf{I} + \mathbf{M}^{-1/2}(\Hessian + \mathbf{S} + \mathbf{F}^T \mathbf{M}^{-1} \mathbf{F}) \mathbf{M}^{-1/2}
               \nonumber\\
             & \ \ \ -\mathrm{i} \mathbf{M}^{-1/2}  (\mathbf{F} + \mathbf{F}^T) \mathbf{M}^{-1/2}.
\end{align}
The formal similarity with the quantum-mechanical random-phase approximation is striking~\cite{MCWEENY92}.

As shown above, the Hamiltonian description of small vibrations yields a
linearized eigenvalue problem of double dimension already from the
start. However, it has the disadvantage of being formulated in terms
of the manifestly gauge dependent canonical momentum
$\mathbf{P}_{\nucidxA}$, leading to the above matrix where the
ingredients $\mathbf{F}$ and $\mathbf{S}$ are gauge dependent too. This disadvantage  can be alleviated by performing the geometric gauge transformation
\begin{equation}
  \begin{split}
  \mathbf{a}'(\mathbf{R}) & =
  \mathbf{a}(\mathbf{R}) - \frac{\partial}{\partial \mathbf{R}} \left( \frac{1}{2} \mathbf{s}^\mathrm T \mathbf{F}(\mathbf{R}_0) \, \mathbf{s}\right)
  \\
  & = \mathbf{a}(\mathbf{R}) - \frac{1}{2} \left( \mathbf{F}(\mathbf{R}_0) + \mathbf{F}(\mathbf{R}_0)^\mathrm T \right) \mathbf{s}
  \end{split}
\end{equation}
where we recall that $\mathbf{s} = \mathbf{R} - \mathbf{R}_0$.
This leaves the quantity defined in Eq.~\eqref{eqSDEF} unchanged (i.e.\ $\mathbf{S}'(\mathbf{R}_0) = \mathbf{S}(\mathbf{R}_0)$), while the first-order derivative
\begin{equation}
  \begin{split}
    F'_{\nucidxB\cartidxB, \nucidxA\cartidxA}(\mathbf{R}_0) & = \left. \frac{\partial a'_{\nucidxB\cartidxB}(\mathbf{R})}{\partial R_{\nucidxA\cartidxA}} \right|_{\mathbf{R}=\mathbf{R}_0}
    \\
    & = \tfrac{1}{2} F_{\nucidxB\cartidxB, \nucidxA\cartidxA}(\mathbf{R}_0) - \tfrac{1}{2} F_{\nucidxA\cartidxA, \nucidxB\cartidxB}(\mathbf{R}_0) \\ &= \tfrac{1}{2} \TotRotComp_{\nucidxB\cartidxB,\nucidxA\cartidxA}(\mathbf{R}_0)
  \end{split}
\end{equation}
has now been transformed into half the Berry curvature defined in Eq.~\eqref{eqTotRotDef}. After this geometric gauge transformation the Hamiltonian matrix takes the form
\begin{equation}
  \begin{split}
    \mathcal{H}^{\prime[2]}_{\text{mat}} (\mathbf{R}_0) & = 
    \begin{pmatrix}
      \Hessian + \mathbf{S} + \tfrac{1}{4} \TotRot^\mathrm T \mathbf{M}^{-1} \TotRot & \tfrac{1}{2} \TotRot^\mathrm T \mathbf{M}^{-1}\\
      \tfrac{1}{2} \mathbf{M}^{-1} \TotRot & \mathbf{M}^{-1}
    \end{pmatrix},
  \end{split}
\end{equation}
where $\mathbf{S}$ is the remaining gauge dependent contribution.

In what follows, we neglect $\mathbf{S}$. Noting that the original momentum $\hat{\boldsymbol{\kappa}}$ is now gauge transformed into some $\hat{\boldsymbol{\kappa}}'$, the eigenvalue equation is
\begin{equation}
  \begin{split}
    \begin{pmatrix}
        \mathrm i\omega \hat{\DisplVec} \\ \mathrm i\omega \hat{\boldsymbol{\kappa}}'
      \end{pmatrix}
    & = \begin{pmatrix} \mathbf{0} & \mathbf{I} \\ -\mathbf{I} & \mathbf{0} \end{pmatrix} \mathcal{H}^{\prime[2]}_{\text{mat}}  \begin{pmatrix}
     \hat{\DisplVec} \\ \hat{\boldsymbol{\kappa}}'
   \end{pmatrix}
  \\
  & = \begin{pmatrix}
    \tfrac{1}{2} \mathbf{M}^{-1} \TotRot \hat{\DisplVec} + \mathbf{M}^{-1} \hat{\boldsymbol{\kappa}}' \\
    -(\Hessian + \tfrac{1}{4} \TotRot^\mathrm T \mathbf{M}^{-1} \TotRot) \hat{\DisplVec} -  \tfrac{1}{2} \TotRot^\mathrm T \mathbf{M}^{-1} \hat{\boldsymbol{\kappa}}'
  \end{pmatrix}.
  \end{split}
\end{equation}
The upper half of this system yields $\hat{\boldsymbol{\kappa}}'   = (\mathrm i\omega \mathbf{M} - \tfrac{1}{2} \TotRot) \hat{\DisplVec}$. Inserting this into the lower half, and using the anti-symmetry $\TotRot^T = -\TotRot$, yields precisely the quadratic eigenvalue problem in Eq.~\eqref{eqQEP}.

\subsection{Orthogonality properties}
\label{secHamOrthoProp}

Consider now multiple solutions $\hat{\DisplVec}_k,\hat{\boldsymbol{\kappa}}_k,\omega_k$ to the above eigenvalue problem. Projecting Eq.~\eqref{eqHamLinearizedJasMetric} from the left by a different solution yields
\begin{equation}
  -\mathrm i\omega_k \begin{pmatrix} \hat{\DisplVec}_l \\ \hat{\boldsymbol{\kappa}}_l \end{pmatrix}^{\dagger} \begin{pmatrix} \mathbf{0} & \mathbf{I} \\ -\mathbf{I} & \mathbf{0} \end{pmatrix} \begin{pmatrix} \hat{\DisplVec}_k \\ \hat{\boldsymbol{\kappa}}_k \end{pmatrix} = \begin{pmatrix} \hat{\DisplVec}_l \\ \hat{\boldsymbol{\kappa}}_l \end{pmatrix}^{\dagger} \mathcal{H}^{[2]}_{\text{mat}} \begin{pmatrix} \hat{\DisplVec}_k \\ \hat{\boldsymbol{\kappa}}_k \end{pmatrix}
\end{equation}
with complex conjugate
\begin{equation}
  -\mathrm i\omega_k \begin{pmatrix} \hat{\DisplVec}_k \\ \hat{\boldsymbol{\kappa}}_k \end{pmatrix}^{\dagger} \begin{pmatrix} \mathbf{0} & \mathbf{I} \\ -\mathbf{I} & \mathbf{0} \end{pmatrix} \begin{pmatrix} \hat{\DisplVec}_l \\ \hat{\boldsymbol{\kappa}}_l \end{pmatrix} = \begin{pmatrix} \hat{\DisplVec}_k \\ \hat{\boldsymbol{\kappa}}_k \end{pmatrix}^{\dagger} \mathcal{H}^{[2]}_{\text{mat}} \begin{pmatrix} \hat{\DisplVec}_l \\ \hat{\boldsymbol{\kappa}}_l \end{pmatrix}.
\end{equation}
Taking the difference between the first equation with $k=1$ and $l=2$ and the second equation  with $k=2$ and $l=1$, we obtain
\begin{equation}
  \label{eqHamOrthoDagger}
  (\omega_1-\omega_2) \begin{pmatrix} \hat{\DisplVec}_2 \\ \hat{\boldsymbol{\kappa}}_2 \end{pmatrix}^{\dagger} \begin{pmatrix} \mathbf{0} & \mathbf{I} \\ -\mathbf{I} & \mathbf{0} \end{pmatrix} \begin{pmatrix} \hat{\DisplVec}_1 \\ \hat{\boldsymbol{\kappa}}_1 \end{pmatrix} = 0.
\end{equation}
Repeating the above derivation but with Hermitian conjugates replaced by matrix transposition also yields
\begin{equation}
    \label{eqHamOrthoTranspose}
  (\omega_1+\omega_2) \begin{pmatrix} \hat{\DisplVec}_2 \\ \hat{\boldsymbol{\kappa}}_2 \end{pmatrix}^\mathrm T \begin{pmatrix} \mathbf{0} & \mathbf{I} \\ -\mathbf{I} & \mathbf{0} \end{pmatrix} \begin{pmatrix} \hat{\DisplVec}_1 \\ \hat{\boldsymbol{\kappa}}_1 \end{pmatrix} = 0.
\end{equation}
Hence, when the frequencies satisfy $\omega_1^2 \neq \omega_2^2$, we obtain two simultaneous orthogonality properties. When the frequencies satisfy either $\omega_1 = \omega_2 \neq 0$ or $\omega_1 = -\omega_2 \neq 0$, only one of the properties is informative.

\subsection{Quantization and interpretation of frequencies as excitation energies}

The classical theory of vibrations derived above reduces to the
standard one in the limit $\TotRot \to \mathbf{0}$ of vanishing
velocity-dependent forces. Even though the calculated frequencies are
derived as dynamical resonance frequencies, they are usually
interpreted as excitation energies. In the case $\TotRot = \mathbf{0}$,
this interpretation is justified by the analogy between a \emph{single} classical and
quantum harmonic oscillator. The case of $3N$ coupled harmonic
oscillators may be reduced to that of a single oscillator by exploiting the fact 
that the $3N$ normal coordinates are orthogonal and
diagonalize $\smash{\mathcal{H}^{[2]}_{\text{mat}}}$, resulting in $3N$
decoupled oscillators~\cite{ATKINS_1997,GRAYBEAL_1988}. When $\TotRot \neq \mathbf{0}$, the orthogonality properties in Sec.~\ref{secHamOrthoProp} are sufficient to achieve such a decoupling, despite the off-diagonal blocks of $\mathcal{H}^{[2]}_{\text{mat}}$ and the lack of orthogonality of the spatial components $\hat{\DisplVec}_k$ discussed in Sec.~\ref{secQEPFreqOrtho}. However, below we follow an alternative approach that circumvents the need for decoupling the oscillators~\cite{QIONGGUI_CTP38_667} and that yields complementary insights.

To this end, consider the quantized analogue of $\mathcal{H}$, obtained by replacing the classical canonical momentum $\mathbf{P}_{\nucidxA}$ by the hermitian operator $\hat{\mathbf{P}}_{\nucidxA} = -\mathrm i\partial/\partial \mathbf{R}_{\nucidxA}$. Collecting all positions and momenta in a single column vector
\begin{equation}
  \hat{\mathbf{z}} = \begin{pmatrix} \mathbf{R} - \mathbf{R}_0 \\ \hat{\mathbf{P}} + \mathbf{a}(\mathbf{R}_0) \end{pmatrix},
\end{equation}
the quantized Hamiltonian becomes
\begin{equation}
  \hat{\mathcal{H}}_{\text{Q}}  = \frac{1}{2} \hat{\mathbf{z}}^{\dagger} \mathcal{H}^{[2]}_{\text{mat}} \hat{\mathbf{z}}.
\end{equation}
With an additional binary index indicating the upper or lower half of $\hat{\mathbf{z}}$, we index the $6N$ components according to  $\hat{z}_{0\nucidxA\cartidxA} = R_{\nucidxA\cartidxA}$ and $\hat{z}_{1\nucidxA\cartidxA} = \hat{P}_{\nucidxA\cartidxA}$. For compactness, we introduce single symbols $a = (\sigma,\nucidxA,\cartidxA)$, $b = (\sigma',\nucidxB,\cartidxB)$, and so on, for such composite indices. Now, defining a matrix $\mathcal J$ with
elements $\mathcal{J}_{ab} =\smash{ -\mathrm{i}[\hat{z}_a,\hat{z}_b]}$, we use the
canonical commutation relations, $[R_a,\hat{P}_b] = \mathrm{i}\delta_{ab}$, to
deduce the matrix form
\begin{equation}
  \mathcal{J} = \begin{pmatrix} \mathbf{0} & \mathbf{I} \\ -\mathbf{I} & \mathbf{0} \end{pmatrix} ,
\end{equation}
which is identical to the matrix that first appeared in Eq.~\eqref{eqHamiltonsEOM} above.

Let us now seek a ladder operator of the form $\hat{X} = \xi_c \hat{z}_c$ (summation
implied over repeated composite indices) and with the property
\begin{equation}
 [\hat{\mathcal{H}}_{\text{Q}}, \hat{X}] = \omega \hat{X}. \label{eqComRel}
\end{equation}
When this operator acts on an energy eigenstate $\ket{0}$, satisfying $\smash{\hat{\mathcal{H}}_{\text{Q}}} \ket{0} = E_0 \ket{0}$, it produces a new eigenstate with an energy shifted by $\omega$:
\begin{equation}
  \hat{\mathcal{H}}_{\text{Q}} \hat{X}\ket{0} = (\hat{X} \hat{\mathcal{H}}_{\text{Q}} + [\hat{\mathcal{H}}_{\text{Q}}, \hat{X}]) \ket{0}
  = (E_0+\omega) \hat{X} \ket{0}.
\end{equation}
Writing out the commutation relation in Eq.\,\eqref{eqComRel} in component form, this relation becomes
\begin{equation}
  \begin{split}
    \omega \xi_e \hat{z}_e & = \frac{1}{2} [\mathcal{H}^{[2]}_{\text{mat};ab} \hat{z}_a \hat{z}_b, \xi_c \hat{z}_c]
    \\
    & = \frac{1}{2} \mathcal{H}^{[2]}_{\text{mat};ab} \left(\xi_c \hat{z}_a [\hat{z}_b, \hat{z}_c]
    + \xi_c [\hat{z}_a, \hat{z}_c] z_b \right)
    \\
    & = \frac{1}{2} \mathcal{H}^{[2]}_{\text{mat};ab} \left( \xi_c \hat{z}_a \mathrm{i} \mathcal{J}_{bc} +  \xi_c \mathrm i \mathcal{J}_{ac}  \hat{z}_b \right).
  \end{split}
\end{equation}
Renaming dummy summation indices and using the fact that $\smash{\mathcal{H}^{[2]}_{\text{mat}}}$ is a symmetric matrix, we may now verify the matrix eigenvalue equation
\begin{equation}
  \begin{split}
   \mathrm i \mathcal{H}^{[2]}_{\text{mat}} \mathcal{J} \boldsymbol{\xi}
    & = \omega \boldsymbol{\xi}.
  \end{split}
\end{equation}
After a change of variables $\boldsymbol{\xi}' = \mathcal{J} \boldsymbol{\xi}$,  we finally obtain
\begin{equation}
  \begin{split}
    \mathcal{H}^{[2]}_{\text{mat}} \boldsymbol{\xi}'
    & = \mathrm i\omega \mathcal{J} \boldsymbol{\xi}'.
  \end{split}
\end{equation}
This eigenvalue equation is formally identical to the complex conjugate of Eq.~\eqref{eqHamLinearizedJasMetric} and we have
\begin{equation}
  \boldsymbol{\xi}' \propto \begin{pmatrix} \hat{\DisplVec}^* \\ \hat{\boldsymbol{\kappa}}^* \end{pmatrix},
\end{equation}
where the quantities on the right-hand side are classical Fourier coefficients rather than quantum-mechanical operators. \textit{In summary, the same mathematical eigenvalue problem yields both (a) classical modes with associated resonance frequencies and (b) quantum-mechanical ladder operators with associated energy shifts.} This identification justifies the interpretation of frequencies as  excitation energies.

For two different ladder operators, $\hat{X}_k$ and $\hat{X}_l$,
\begin{equation}
  \begin{split}
    [\hat{X}_k,\hat{X}_l] & = \xi_{k;a} [\hat{z}_a,\hat{z}_b] \xi_{l;b}
    = \mathcal{J}_{ac} \xi'_{k;c} [\hat{z}_a,\hat{z}_b] \mathcal{J}_{db} \xi'_{l;d}
    \\
    & = \mathcal{J}_{ac} \xi'_{k;c} \mathrm i\mathcal{J}_{ab} \mathcal{J}_{db} \xi'_{l;d}
    = -\mathrm i \xi'_{k;c} \mathcal{J}_{cd} \xi'_{l;d}
  \end{split}
\end{equation}
which according to the orthogonality property in Eq.~\eqref{eqHamOrthoTranspose} vanishes if $\omega_k \neq -\omega_l$. Hence, ladder operators with different absolute frequencies $|\omega_k|\neq |\omega_l|$ commute. When $\omega_k = -\omega_l>0$, we have $\smash{\hat{X}_k = \hat{X}_l^{\dagger}}$ and $\smash{\hat{X}_l^{\dagger} \hat{X}_l}$ is (up to normalization) the number operator for the $k$th mode.
  
\section{Results} \label{secResults}

We illustrate the above theory of small oscillations for a few molecular systems subject to strong a magnetic field. All calculations were performed at the Hartree--Fock level using {\sc London}~\cite{LondonProgram,TELLGREN_JCP129_154114}. In particular, its functionality for analytical geometrical gradients~\cite{TELLGREN_PCCP14_9492} and the recent implementation of the Berry-curvature tensor~\cite{CULPITT_JCP155_024104,CULPITT_JCP156_044121} was relied on. Standard basis-set names are prefixed by `L' to indicate that London gauge factors~\cite{LONDON_JPR8_397} are used and by `u' to indicate decontraction of primitive Gaussian functions. For a strong magnetic field, decontraction is useful since most basis sets were originally constructed for zero-field situations. Hessians were computed by numerical differentiation of analytical gradients~\cite{TELLGREN_PCCP14_9492}. For each system, the optimal geometry, including the optimal orientation relative to the field, was determined as the minimum of the potential-energy surface for a given field strength. The quadratic eigenvalue problem was solved using the function \texttt{polyeig} in the {\sc Matlab} package.

\subsection{Singlet H$_2$}

\begin{figure}
  \includegraphics[width=0.95\linewidth]{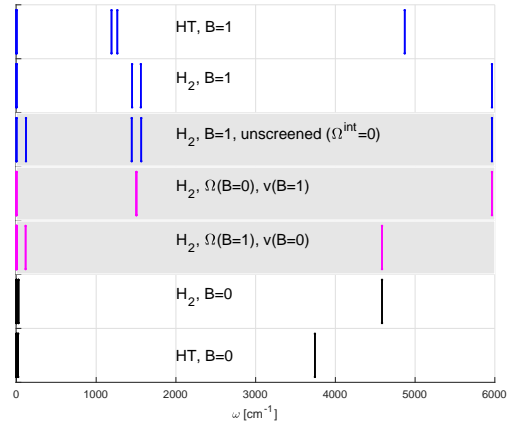}
  \caption{\label{figH2spec} Spectrum of the hydrogen molecule in the singlet electronic state and the minimum geometry parallel to the magnetic field of strength $B=0$ and $B=B_0$. The top and bottom spectra show the tritium isotopologue HT, whereas the other spectra are for H$_2$. For comparison, the shaded region also displays the spectra produced with mixed data---with $\TotRot$ and $\PES$ computed at different field strengths. The vertical axis is arbitrary and for visualization purposes only.}
\end{figure}

At the RHF/Lu-aug-cc-pVTZ level of theory, the optimal bond distance at $B=0$ is $R=1.388$\,bohr. By contrast, at $B=B_0$, the optimal bond distance is $R=1.219$\,bohr, with the bond axis parallel to the magnetic field. In Fig.\,\ref{figH2spec}, the spectrum of the hydrogen molecule at these geometries is shown. In the absence of a field, there are translational modes with $\omega=0$, rotational modes with $\omega=24.3$\,cm$^{-1}$, and a vibrational mode with $\omega=4585$\,cm$^{-1}$. At $B=B_0$, the translational modes with $\omega=0$ remain, but the rotational modes have split into two distinct frequencies $\omega=1450$\,cm$^{-1}$ and $\omega=1560$\,cm$^{-1}$, while the vibrational mode is substantially stiffer, with $\omega=5964$\,cm$^{-1}$. 

To investigate the relative importance of the velocity-dependent term $\TotRot$ and field-induced change in $\PES$, we computed the spectrum with the Hessian $\mathbf H$  at $B=0$ (at $R=1.388$\,bohr) and $\TotRot$ at $B=B_0$ (at $R=1.219$\,bohr). We also interchanged these, so that $\mathbf H$ was obtained at $B=B_0$ (at $R=1.219$~bohr) and $\TotRot=\mathbf 0$ as it is at $B=0$. Another variant we considered was to use $\mathbf H$ at $B=B_0$ with the bare Lorentz force, but neglecting the Berry screening force ($\BerryCurv=\mathbf{0}$).

\begin{figure}
  \includegraphics[width=0.95\linewidth]{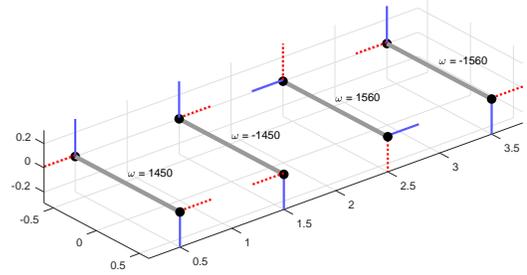}
  \caption{\label{figH2rot} Four rotational eigenvectors $\hat{\DisplVec}(\omega)$ for the H$_2$ molecule in the singlet state and a magnetic field of $B=B_0$. The bond axis is indicated with a thick black lines, the real component of $\hat{\DisplVec}(\omega)$ with solid blue lines, and the imaginary component with dotted red lines. The dynamical frequency is given in units of cm$^{-1}$ for each mode.  Note that the corresponding four eigenvalues are distinct, but there are only two unique and linearly independent eigenvectors.}
\end{figure}

From these results, shown in the shaded part of Fig.~\ref{figH2spec}, we see that the change in frequencies is almost entirely a result of magnetic-field effects on the potential-energy surface---in particular, the compression of the optimal geometry with the resulting higher curvature captured by the Hessian. The main effect of the external Lorentz force and the Berry curvature is to split the rotational modes, which would otherwise be degenerate at $\omega=\pm 1504$\,cm$^{-1}$. Moreover these degenerate modes could be chosen complex, in which case they satisfy the relation
\begin{equation}
  \hat{\DisplVec}(\pm 1504\ \mathrm{cm}^{-1}) = \hat{\DisplVec}(\mp 1504\ \mathrm{cm}^{-1})^*.
\end{equation}
Due to the degeneracy, the modes are not uniquely determined and we are free to choose the modes as the real and imaginary parts of the complex modes, leading real modes. By contrast, the velocity-dependent forces arising from the true $\TotRot$ at $B=B_0$ break time-reversal symmetry, leading to the situation that
\begin{equation}
  \begin{split}
    \hat{\DisplVec}&(+1560\ \mathrm{cm}^{-1}) = \hat{\DisplVec}(-1450\ \mathrm{cm}^{-1})
        \\
     & = \hat{\DisplVec}(-1560\ \mathrm{cm}^{-1})^* = \hat{\DisplVec}(+1450\ \mathrm{cm}^{-1})^*,
  \end{split}
\end{equation}
which is illustrated in Fig.~\ref{figH2rot}.
Consequently, these rotational modes can no longer be chosen to be real; they always have both real and imaginary components.

The importance of the Berry curvature and the resulting screening force is illustrated in the uppermost shaded area of Fig.~\ref{figH2spec}, showing the unscreened spectrum. The unscreened spectrum is very close to the screened spectrum except for spurious modes at $\omega = \pm 120$\,cm$^{-1}$, which is the cyclotron frequency $q_{\text{tot}} B/m_{\text{tot}}$ for the center of mass. These spurious modes result from the bare Lorentz force and correspond to complex-valued linear combinations of translations in the plane perpendicular to the bond axis.

In more detail, we can order the eigenvectors so that the real-valued $\hat{\DisplVec}_1$ and $\hat{\DisplVec}_2$ are two such translations, both with a vanishing frequency $\omega_1=\omega_2=0$, meaning that $\Hessian \hat{\DisplVec}_1 = \Hessian \hat{\DisplVec}_2 = \mathbf{0}$. The corresponding spurious modes are then $\hat{\DisplVec}_3 = \hat{\DisplVec}_1 + \mathrm i \hat{\DisplVec}_2$ and $\hat{\DisplVec}_4 = \hat{\DisplVec}_3^* = \hat{\DisplVec}_1 - \mathrm i \hat{\DisplVec}_2$, with nonvanishing frequencies $\omega_3=-\omega_4=119.6$\,cm$^{-1}$. Like the real modes $\hat{\DisplVec}_1$ and $\hat{\DisplVec}_2$, they satisfy $\Hessian \hat{\DisplVec}_3 = \Hessian \hat{\DisplVec}_4 = \mathbf{0}$ but, unlike the real modes, they also satisfy
\begin{align}
  (-\mathrm i\omega_3\TotRot - \omega_3^2 \mathbf{M}) \hat{\DisplVec}_3 & = 0, \\
  (-\mathrm i\omega_4\TotRot - \omega_4^2 \mathbf{M}) \hat{\DisplVec}_4 & = 0,
\end{align}
with nonzero frequency. Hence, the complex linear combinations solve two eigenvalue problems simultaneously, while the real valued vectors solve only one.

 As seen in Fig.~\ref{figH2spec},  tritiumation  reduce the rotational and vibrational frequencies. Even though tritiumation breaks inversion symmetry and leads to different gyromagnetic ratios for the two nuclei, the modes remain pure translational, pure rotational, and pure vibrational. For each mode, one of $P_{\mathrm{cm}}$, $P_{\mathrm{rot}}$, or $P_{\mathrm{vib}}$ is close to one and the remaining two are thus close to zero. For the isotopologue HT oriented along the $x$-axis, the translational--rotational coupling matrix of Eq.\,\eqref{eqGmat} is given by
\begin{equation}
  \boldsymbol{\mathcal{G}} =
  \begin{pmatrix}    
         0   &      0     &    0       \\
         0   & -0.609 &         0   \\
         0    &     0  & -0.609   
  \end{pmatrix},
\end{equation}
with vanishing contribution from the Berry curvature. Hence, rotation around the $y$- or $z$-axis, which are perpendicular to both the bond axis and the magnetic field, couples to the center-of-mass translation along the same axis. 
For H$_2$, by contrast, the whole coupling matrix vanishes, $\boldsymbol{\mathcal{G}}=\mathbf{0}$. According to the other quantitative measures, shown for HT in Table~\ref{tabSingletHTmodes}, the modes are to a very high precision either purely translational, rotational, or vibrational.

\begin{table}
  \caption{\label{tabSingletHTmodes} Dynamical frequencies $\omega$ of the singlet HT molecule in a perpendicular field of $B=B_0$. The next three columns show the degree of rigid center-of-mass motion, rigid rotational motion, and vibration (positive and negative frequencies share the same values for this system). The fourth column shows the time-averaged rotational energy.}
  \begin{tabular}{r|lll|l}
\hline \hline
  $\omega$ [cm$^{-1}$] & $P_{\mathrm{cm}}$ & $P_{\mathrm{rot}}$ & $P_{\mathrm{vib}}$ & $\bar{T}_{\text{rot}}/\epsilon^2 \omega^2$ \\ \hline
 $\pm 0.00 $ & 1.000 & 0.000 & 0.000 & 0 \\         
 $\pm 0.00 $ & 1.000 & 0.000 & 0.000 & 0 \\
 $\pm 0.00 $ & 1.000 & 0.000 & 0.000 & 0 \\ 
 $\pm 1192 $ & 0.000 & 1.000 & 0.000 & $2.2\times 10^3$ \\  
 $\pm 1265 $ & 0.000 & 1.000 & 0.000 & $2.2\times 10^3$ \\   
 $\pm 4870 $ & 0.000 & 0.000 & 1.000 & 0 \\ \hline
  \end{tabular}
\end{table}

We finally note that, for both H$_2$ and HT, the Lorentz force on the center of mass is completely canceled by the Berry screening force, according to the magnetic-translational sum rule~\cite{PETERS_JCP157_134108}:
\begin{equation}
  \label{eqH2CMscreen}
  \sum_{\nucidxA \nucidxB} \BerryCurv_{\nucidxA\nucidxB} = q_{\mathrm{tot}} \SkewSymEncode{\mathbf{B}}.
\end{equation}

\begin{figure}
  \includegraphics[width=0.9\linewidth]{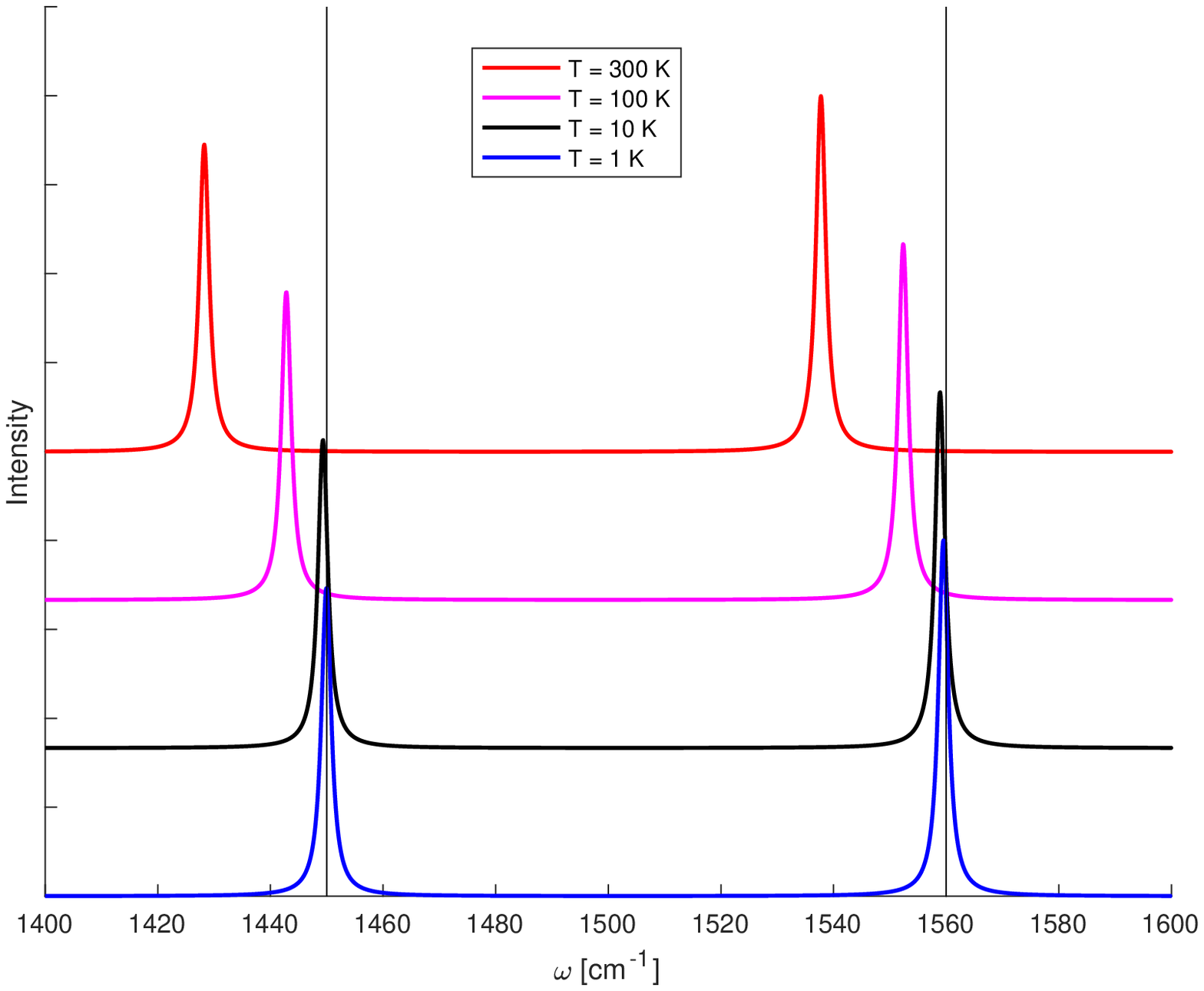}
  \includegraphics[width=0.9\linewidth]{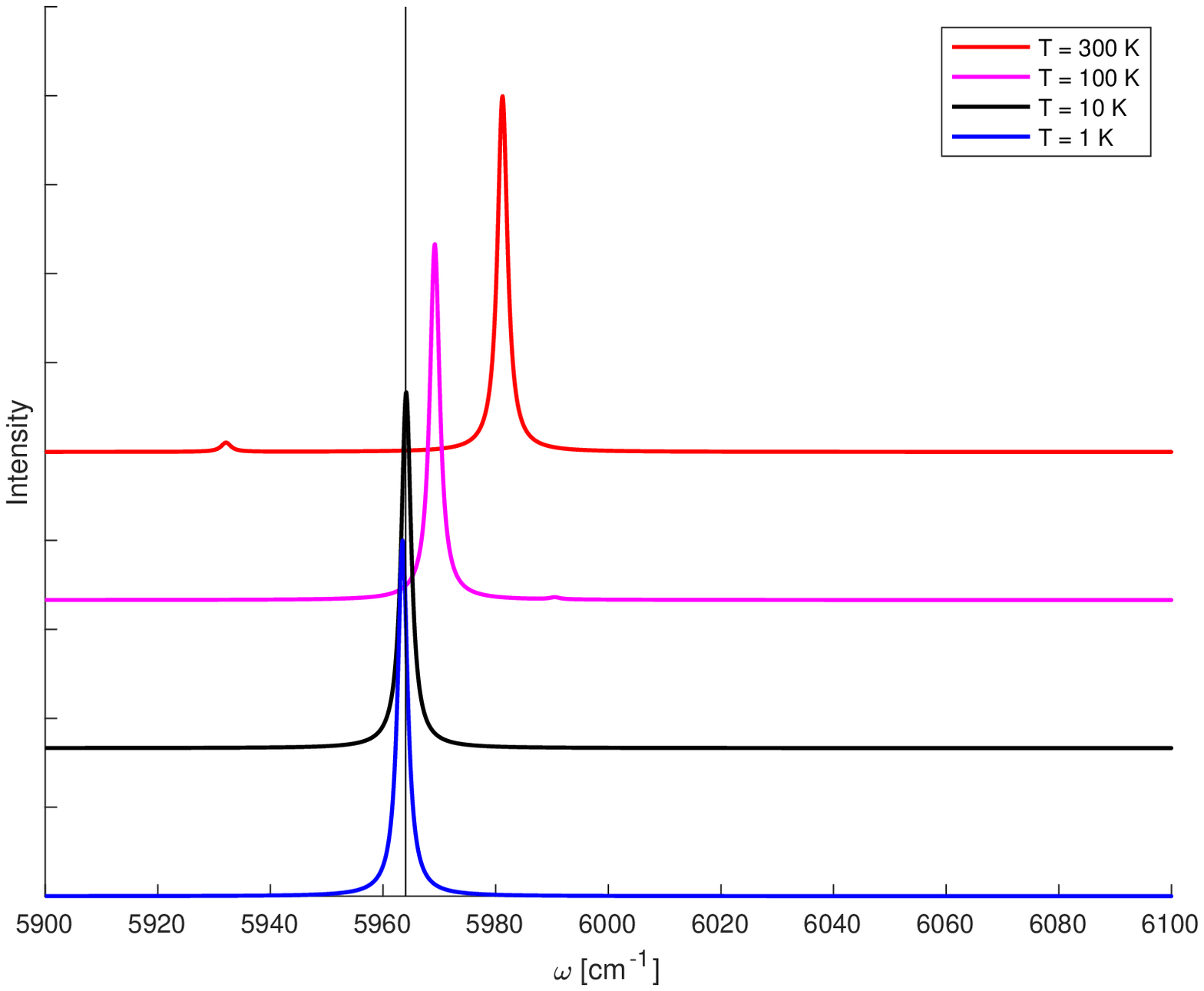}
  \caption{\label{figH2dynspec} Spectrum for singlet H$_2$ at
    $B=B_0$ obtained from Born--Oppenheimer dynamics compared to the lines from the quadratic eigenvalue problem (vertical black lines). The dynamical spectra differ in the initial kinetic energy, which is given in units of kelvin.}
\end{figure}

\begin{figure}
  \includegraphics[width=0.9\linewidth]{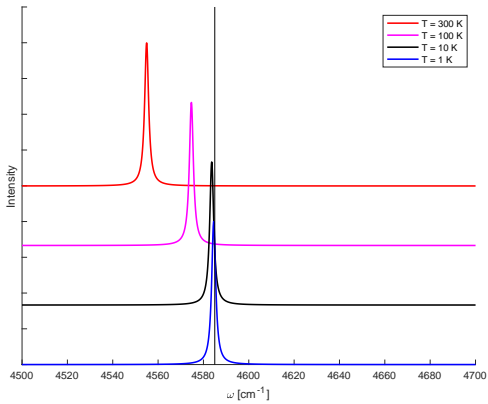}
  \caption{\label{figH2dynspecZeroB} Spectrum near the stretching mode for singlet H$_2$ at vanishing field,
    $B=0$, obtained from Born--Oppenheimer dynamics compared to the lines from the quadratic eigenvalue problem (vertical black lines). The dynamical spectra differ in the initial kinetic energy, which is given in units of kelvin.}
\end{figure}

\subsection{Comparison with dynamical trajectories}

To assess the effects of anharmonicity, we have generated vibrational spectra from Born--Oppenheimer dynamics at different initial conditions for $B=0$ and $B=B_0$, with inclusion of the Lorentz force and the Berry screening force. As in previous work~\cite{PETERS_JCP155_024105}, we use the auxiliary-coordinates-and-momenta propagator, a time step of 1\,fs, and a total simulation time of 200\,ps. The simulations started from the equilibrium geometry at the given field strength with initial kinetic energy, in units of kelvin, 0.01, 0.1, 1, 10, 100, and 300\,K. For the simulations at $B=0$, we do not sample the the rotational modes. The resulting spectra for the higher kinetic energies are shown in Fig.\,\ref{figH2dynspec}; the lowest kinetic energies yield spectra visually indistinguishable from $T=1$\,K.

At a sufficiently low kinetic energy, the motion is expected to be near harmonic, whereas a higher initial kinetic energy will probe anharmonic effects. Our results confirm this picture, with the dynamical spectrum coinciding with the harmonic approximation up to about 1\,K. Interestingly, the anharmonic effects on the stretching vibration  produce a red shift at $B=0$ (see Fig.~\ref{figH2dynspecZeroB}) but a blue shift at $B=B_0$. While the former effect is well known, the latter can be understood from the observation that the covalent hydrogen $\sigma_{\mathrm{g}}^2$ bond becomes stiffer as the angle between the magnetic field and the bond axis increases. Since higher temperatures allow the molecule to explore these parts of the potential-energy surface, the averaged frequency of the stretching mode increases.

\subsection{Triplet H$_2$}

Although the triplet state of H$_2$ is unbound in the absence of a magnetic field, it becomes bound in a sufficiently strong magnetic field~\cite{KUBO_JPCA111_5572,LANGE_S337_327}, with a preferred perpendicular orientation in the field. The underlying mechanism for this \emph{perpendicular paramagnetic bonding} is that anti-bonding orbitals develop an orientation-dependent angular momentum that is stabilized by the orbital Zeeman effect~\cite{LANGE_S337_327, TELLGREN_PCCP14_9492,STOPKOWICZ_JCP143_074110,AUSTAD_PCCP22_23502}. At the UHF/Lu-aug-cc-pVTZ level, the optimal bond distance at $B=B_0$ is $R=2.709$\,bohr, at the preferred perpendicular field orientation. At this field strength, the $M_\text{S} = -1$ component of the triplet is the ground state of H$_2$.

\begin{figure}
  \includegraphics[width=0.95\linewidth]{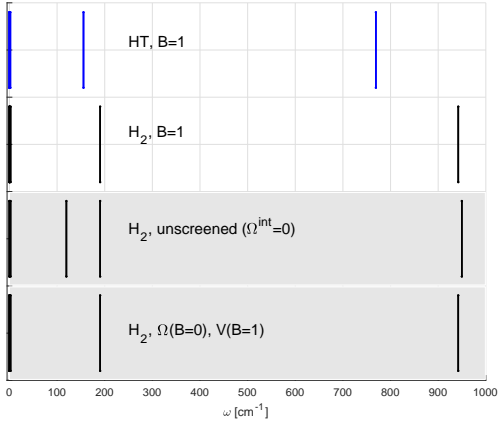}
  \caption{\label{figH2TripletSpec} Spectrum of the hydrogen molecule in the triplet electronic state and the minimum geometry perpendicular to the magnetic field of strength $B=B_0$. The top spectrum shows the tritium isotopologue HT, whereas the other spectra are for H$_2$. For comparison, the shaded region also displays the spectra produced by zeroing out either the full $\TotRot=\mathbf{0}$ or the Berry curvature contribution $\BerryCurv=\mathbf{0}$. The vertical axis is arbitrary and for visualization purposes only.}
\end{figure}

Spectra of triplet H$_2$ and HT are shown in Fig.~\ref{figH2TripletSpec}. As expected, the H$_2$ stretching mode of $942$\,cm$^{-1}$ at $B=B_0$ is much lower than the corresponding frequency of singlet H$_2$. Unlike in the singlet H$_2$, a rotational mode appears at $\omega = 2.4$\,cm$^{-1}$; it corresponds to the barrier-free rotation about the field axis. 
The hindered rotation in the plane spanned by the molecular axis and the magnetic field occurs at at $\omega=\pm 190$\,cm$^{-1}$; it is not degenerate and the velocity-dependent force does not split this mode. As a result, the spectra obtained with screened and unscreened Lorentz forces are very similar; see the lower shaded area in Fig.\,\ref{figH2TripletSpec}. However, as seen in the upper shaded area in Fig.~\ref{figH2TripletSpec}, omission of the Berry screening force leads to a spurious center-of-mass cyclotron mode at $\omega=\pm 120$\,cm$^{-1}$, since the system then behaves like particle of charge $+2$ in a magnetic field.

\begin{table}
  \caption{\label{tabTripletH2HTmodes} Dynamical frequencies $\omega$ of triplet H$_2$ and HT in a perpendicular field of $B=B_0$. The next three columns show the degree of rigid center-of-mass motion, rigid rotational motion, and vibration (positive and negative frequencies share the same values for this system). The fourth column shows the time-averaged rotational energy.}
  \begin{tabular}{r|lll|l}
    \hline \hline
                 &           & H$_2$ &   & \\ \hline
 $\pm 0 $ & 1.000 & 0.000 & 0.000 & 0 \\     
 $\pm 0 $ & 1.000 & 0.000 & -0.000 & 0 \\  
 $\pm 0 $ & 1.000 & 0.000 & 0.000 & 0 \\   
 $\pm 2.4 $ & 0.000 & 1.000 & 0.000 & $1.8\times 10^3$ \\   
 $\pm 190 $ & 0.000 & 1.000 & -0.000 & $1.8\times 10^3$ \\  
  $\pm 942 $ & 0.000 & 0.000 & 1.000 & 0.30 \\      
    \hline \hline
                 &           & HT &   & \\ \hline
 $\pm 0 $ & 1.000 & 0.000 & 0.000 & 0 \\ 
 $\pm 0 $ & 1.000 & 0.000 & 0.000 & 0 \\
 $\pm 0 $ & 1.000 & 0.000 & 0.000 & 0 \\ 
 $\pm 1.9 $ & 0.000 & 1.000 & 0.000 & $2.2\times 10^3$ \\    
 $\pm 155 $ & 0.000 & 1.000 & 0.000 & $2.2\times 10^3$  \\    
 $\pm 769 $ & 0.000 & 0.000 & 1.000 & 0.24 \\     \hline
  \end{tabular}
\end{table}

As seen in Table~\ref{tabTripletH2HTmodes}, all modes in triplet H$_2$ are to high precision pure translations, rotations, and vibrations according to our quantification using $P_{\text{cm}}$, $P_{\text{vib}}$, and $P_{\text{vib}}$. However, while the zero-frequency modes have a vanishing time average $\bar{T}/\epsilon^2\omega^2 = 0$ (this quantity can be calculated without division by zero), the stretching mode has a small rotational energy $\bar{T}/\epsilon^2\omega^2 = 0.3$ that is not numerical noise. This amount should be compared to $\bar{T}/\epsilon^2\omega^2 = 1.8\times 10^3$ for the rotational modes. 

For the isotopologue HT oriented along the $x$-axis and the magnetic field along the $z$-axis, the translational-rotational coupling matrix is given by
\begin{equation}
  \boldsymbol{\mathcal{G}} =
  \begin{pmatrix}    
         0   &      0     &    1.354       \\
         0   &     0 &         0   \\
         0    &     0  &    0  
  \end{pmatrix},
\end{equation}
with vanishing contribution from the Berry curvature. Hence, rotation about the $z$-axis couples to center-of-mass translation along the $x$-axis. For H$_2$, the coupling matrix vanishes, $\boldsymbol{\mathcal{G}}=\mathbf{0}$. For both H$_2$ and HT, the bare Lorentz force on the center of mass is  canceled by the Berry screening force; see Eq.~\eqref{eqH2CMscreen}.

\subsection{Singlet HCN}

Hydrogen cyanide has been investigated at the RHF/Lu-cc-pVDZ level. In a strong field of $B=0.3B_0$, HCN adopts a linear geometry perpendicular to the field, with a C--H and C--N bond distances of 1.964 and 2.153~bohr, respectively. All 18 dynamical frequencies are given in Table~\ref{tabHCNmodes} along with the degree of translational, rotational, and vibrational motion. 

For this system, numerical errors are noticeable in our finite-difference procedure for computing the Hessian, probably because the energy and forces respond very differently to vibrational and rotational modes, respectively. Our numerical Hessian has a nonsymmetric component on the order of $ \|\Hessian - \Hessian^\mathrm T\|_\mathrm F = 1.5\times 10^{-6}\ \text{au} = 0.3\ \text{cm}^{-1}$. Additionally, numerical errors manifest themselves in that some frequencies with absolute value below $0.3\ \text{cm}^{-1}$ are imaginary. For this reason we report frequencies $|\omega| < 0.3\ \text{cm}^{-1}$ as 0. 

\begin{figure}
  \begin{center}
  \includegraphics[width=0.98\linewidth]{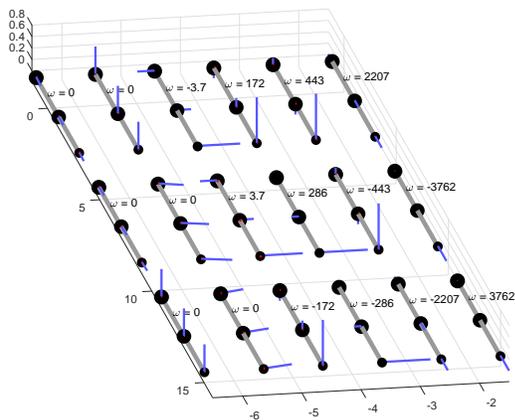}
  \caption{\label{figHCNmodes} Oscillatory modes of the HCN molecule oriented perpendicular to a field of $B=0.3B_0$. The real part of each mode is indicated with blue lines. Only the mode with $\omega=3.7$\,cm$^{-1}$ has a (barely) visible imaginary part indicated in red. The length scale for these lines is arbitrary; however, the relative lengths within and between modes are meaningful. The dynamical frequency $\omega$ in cm$^{-1}$ is also given for each mode.}
  \end{center}
\end{figure}

The  modes listed in the table are  visualized in Fig.\,\ref{figHCNmodes}. Several modes mixed rather than pure translations, rotations, or vibrations---for example, the rotational mode with $\omega = \pm 3.7$\,cm$^{-1}$, corresponding to rotation about the field axis, includes 10.5\% center-of-mass motion. (This mode appears particularly sensitive to numerical errors. In initial calculations, with a less accurate Hessian with $\|\Hessian - \Hessian^\mathrm T\|_{\mathrm{F}} = 2.8\,\text{cm}^{-1}$, we got a frequency of 7.1~cm$^{-1}$ for this mode.) Moreover, the rotational mode at $\omega=172$~cm$^{-1}$, which corresponds to rotation around an axis perpendicular to both the magnetic field and molecular axis, has 5.7\% vibrational character. Similarly, the vibrational mode with $\omega=443$~cm$^{-1}$, corresponding to a change in the bond angle by displacing atoms along the magnetic field axis, has 5.7\% rotational character.

As for singlet and triplet H$_2$, the Lorentz force on the center of mass is completely canceled by the Berry screening force (see Eq.~\eqref{eqH2CMscreen}). The translational--rotational coupling matrix has contributions both from the external magnetic field and the Berry curvature. With the molecular axis aligned to the $x$-axis and the magnetic field to the $z$-axis, it is obtained from these two contributions as
\begin{equation}
  \boldsymbol{\mathcal{G}} =
  \begin{pmatrix}    
         0   &      0     &    0.451       \\
         0   &     0 &         0   \\
         0    &     0  &    0  
   \end{pmatrix}
       +
  \begin{pmatrix}    
         0   &      0     &    -0.337       \\
         0   &     0 &         0   \\
         0    &     0  &    0  
  \end{pmatrix}.
\end{equation}
This results in a coupling between rigid rotations about the magnetic field ($z$-axis) and center-of-mass translations along the molecular axis ($x$-axis). This coupling is seen in the mode with $\omega = \pm 0$, which is predominantly translation along the molecular axis but with 0.4\% rotational character. \newline
\begin{table}
  \caption{\label{tabHCNmodes} Dynamical frequencies $\omega$ of the HCN molecule in a perpendicular field of $B=0.3B_0$. The next three columns show the degree of rigid center-of-mass motion, rigid rotational motion, and vibration (positive and negative frequencies share the same values for this system). The fourth column shows the time-averaged rotational energy.}
  \begin{tabular}{r|lll|c}
\hline \hline
  $\omega$ [cm$^{-1}$] & $P_{\mathrm{cm}}$ & $P_{\mathrm{rot}}$ & $P_{\mathrm{vib}}$ & $\bar{T}_{\text{rot}}/\epsilon^2 \omega^2$ \\ \hline
 $\pm 0$ & 1.000 & 0.000 & 0.000 & 0 \\ 
 $\pm0$ & 1.000 & 0.000 & 0.000 & 0 \\ 
 $\pm 0$ & 0.996 & 0.004 & 0.000 & 58 \\     
 $\pm 3.7 $ & 0.105 & 0.895 & 0.000 & $6.3 \times 10^3$ \\  
 $\pm 172 $ & 0.000 & 0.943 & 0.057 & $3.5 \times 10^3$ \\  
 $\pm 286 $ & 0.000 & 0.000 & 1.000 & $4.5\times 10^{-5}$ \\  
 $\pm 443 $ & 0.000 & 0.057 & 0.943 & 190 \\       
 $\pm 2207 $ & 0.000 & 0.000 & 1.000 & $1.3\times 10^{-3}$ \\  
 $\pm 3762 $ & 0.000 & 0.000 & 1.000 & $1.2\times 10^{-2}$ \\      \hline
  \end{tabular}

\end{table}

\section{Conclusions}

We have developed the theory of small oscillations in molecules subject to velocity-dependent forces. Specifically, when the velocity-dependent forces are the Lorentz forces in an external magnetic field and the Berry screening force~\cite{YIN_JCP100_8125,CERESOLI_PRB75_161101,CULPITT_JCP155_024104}, several novel effects arise. For example, the internal motion can no longer be considered in isolation and there is coupling between rigid translations, rigid rotations, and pure vibrations. Also, rotational modes are split when the clockwise and counterclockwise rotation are no longer symmetry equivalent but subject to different magnetic forces.

In mathematical terms, the dynamical frequencies are no longer obtained from a standard eigenvalue problem involving the mass matrix and the geometric Hessian. Instead, the frequencies are obtained from a quadratic eigenvalue problem, with a more complicated structure of eigenvectors than a standard eigenvalue problem. We also discussed \emph{linearization} -- that is, the conversion of the quadratic eigenvalue problem to a standard eigenvalue problem of double dimension. The most interesting linearization is obtained from Hamilton's equation of motion. Although this work is focused on the classical dynamics of nuclei, the Hamiltonian formulation admits a close analogy with the quantum-mechanical treatment of vibrations, allowing us to verify that the classical frequencies can be interpreted as excitation energies.

The developed theory has been illustrated by numerical results for the hydrogen molecule and hydrogen cyanide in a strong magnetic field. Although effect of the velocity-dependent forces is in general complex, its main effect is to split rotational modes. By contrast, the effect of the magnetic field via the potential-energy surface is to blue shift stretching frequencies and introduce hindered rotations. Whereas harmonic stretching frequencies are typically blue shifted relative to the (true) anharmonic frequencies in the absence of the magnetic field, we observe the opposite effect for singlet H$_2$ in a strong magnetic field, for which the anharmonic stretching frequency is higher than the corresponding harmonic frequency.

Coupling of translational, rotational, and vibrational motions is also seen in our numerical results, although the magnitude of the coupling is small. This is likely due to the relatively simple structure of the coupling matrix $\boldsymbol{\mathcal{G}}$ in the above linear molecules. Because our quantification of the coupling is energy based, the small coupling could also be due to modes of different type having kinetic energies of different orders of magnitude. Larger molecules with lower symmetry may display such couplings in a larger percentage of the modes.

\section*{Acknowledgments}

This work was supported by the Research Council of Norway through the ``Magnetic Chemistry'' Grant No.\,287950 and CoE Hylleraas Centre for Molecular Sciences Grant No.\,262695. The work also received support from the UNINETT Sigma2, the National Infrastructure for High Performance Computing and Data Storage, through a grant of computer time (Grant No.\,NN4654K).

\end{document}